\documentclass[%
 reprint,
 superscriptaddress,
preprintnumbers,
 amsmath,amssymb,
 aps,prx
]{revtex4-2}
\usepackage[section]{placeins}
\usepackage{amsfonts}
\usepackage{graphicx}
\usepackage{color}
\usepackage[colorlinks=true, pdfstartview=FitV, linkcolor=red, citecolor=blue, urlcolor=blue]{hyperref}
\usepackage{tcolorbox}
\usepackage{tikz}
\usepackage{tikz-cd}
\usetikzlibrary{decorations.pathmorphing}
\tikzset{snake it/.style={decorate, decoration=snake}}
\usetikzlibrary{calc}
\usepackage{tikz-3dplot} 
\tdplotsetmaincoords{70}{120}  
\tdplotsetrotatedcoords{0}{0}{0} 
\usepackage{tikz}
\pgfdeclarelayer{bg}
\pgfsetlayers{bg,main}
\usepackage{algpseudocode,algorithm}

\def\rev#1{{\color{black}#1}} 
\def\revv#1{{\color{black}#1}}

\begin{document}

\title{Bulk and boundary entanglement transitions in the projective gauge-Higgs model}

\author{Hiroki Sukeno}
\email{hiroki.sukeno@gmail.com}
\affiliation{C.N. Yang Institute for Theoretical Physics $\&$ Department of Physics and Astronomy, State University of New York at Stony Brook, Stony Brook, NY 11794-3840
}

\author{Kazuki Ikeda}
\email{kazuki.ikeda@stonybrook.edu}
\affiliation{Center for Nuclear Theory, Department of Physics and Astronomy, Stony Brook University, Stony Brook, New York 11794-3800, USA}
\affiliation{Department of Physics, University of Massachusetts Boston, Boston, MA 02125, USA}

\author{Tzu-Chieh Wei}
\email{tzu-chieh.wei@stonybrook.edu}
\affiliation{C.N. Yang Institute for Theoretical Physics $\&$ Department of Physics and Astronomy, State University of New York at Stony Brook, Stony Brook, NY 11794-3840
}

\date{\today}

\begin{abstract}
In quantum many-body spin systems, the interplay between the entangling effect of multi-qubit Pauli measurements and the disentangling effect of single-qubit Pauli measurements may give rise to two competing effects. By introducing a randomized measurement pattern with such bases, a phase transition can be induced by altering the ratio between them. In this work, we numerically investigate a measurement-based model associated with the $(2+1)$d $\mathbb{Z}_2$ Fradkin-Shenker Hamiltonian that encompasses the deconfining, confining, and Higgs phases. We determine the phase diagram in our measurement-only model by employing entanglement measures. 
For the bulk topological order, we use the topological entanglement entropy. 
We also use the mutual information between separated boundary regions to diagnose the boundary phase transition associated with the Higgs or the bulk SPT phase.
We observe the structural similarity between our phase diagram and the one in the standard quantum Hamiltonian formulation of the Fradkin-Shenker model with the open rough boundary. 
First, a deconfining phase is detected by nonzero and constant topological entanglement entropy. 
Second, we find a (boundary) phase transition curve separating the Higgs=SPT phase from the rest.
In certain limits, the topological phase transitions reside at the critical point of the formation of giant homological cycles in the bulk 3d spacetime lattice, as well as the bond percolation threshold of the boundary 2d spacetime lattice when it is effectively decoupled from the bulk. 
Additionally, there are analogous mixed-phase properties at a certain region of the phase diagram, emerging from how we terminate the measurement-based procedure.
Our findings pave an alternative pathway to study the physics of Higgs=SPT phases on quantum devices in the near future. 
\end{abstract}

\maketitle

\section{Introduction}

Entanglement plays a central role in quantum physics.
There are numerous examples of quantum many-body systems that exhibit intricate patterns of entanglement in their ground states.
An outstanding example of such entanglement properties is the long-range entanglement (LRE) \cite{bravyi2006lieb,chen2010local}---also known as the topological order---possessed by Kitaev's toric code \cite{kitaev2003fault}, the string-net models \cite{levin2005string}, and so on. 
In the absence of LRE, a state is said to have short-range entanglement (SRE). 
Such a state may have a nontrivial entanglement pattern if one imposes global symmetries, giving rise to the notion of symmetry-protected topological (SPT) order~\cite{gu2009tensor,pollmann2012symmetry, pollmann2010entanglement, chen2011classification, chen2011two, chen2011complete, schuch2011classifying,chen2013symmetry}. 
Classification of entanglement in the low-energy regime of quantum many-body systems has been one of the major subjects in condensed matter physics over a decade~\cite{wen2017colloquium}.

In recent years, there has been growing interest in the study of phase transitions in entanglement under evolution involving both unitary operators and non-unitary operators such as measurements~\cite{bardarson2012unbounded, serbyn2013universal, nahum2017quantum, von2018operator}. 
The interplay between multi-qubit entangling unitary gates and single-qubit measurements exhibits competing effects on entanglement and can lead to phase transitions when the rate of projective measurement is varied, which is known as measurement-induced phase transitions (MIPT)~\cite{li2018quantum, li2019measurement, chan2019unitary, tang2020measurement,alberton2021entanglement}.
Some studies have focused on special classes of unitary gates, such as the Clifford gates or Majorana fermions, which can be efficiently simulated classically in combination with Pauli measurements due to the Gottesman-Knill theorem~\cite{li2018quantum, li2019measurement, chan2019unitary, gottesman1996class, aaronson2004improved,2021NatPh..17..342L, PRXQuantum.2.040319, 2020PhRvB.102a4315T, lavasani2021topological, sierant2022measurement,
zhu2023structured, zabalo2022operator, sang2021entanglement, PhysRevResearch.3.023200, li2021statistical, lunt2021measurement, morral2023detecting, zabalo2020critical, liu2022measurement, li2021conformal}. 
These investigations have provided insights into the critical behavior of MIPT with fairly large system sizes. 
Other works have mapped MIPT circuits, including both Clifford and random (Haar) unitary circuits, to classical statistical models, such as the percolation model~\cite{2019PhRvX...9c1009S, lavasani2021topological, sierant2022measurement, PhysRevResearch.2.023288, sierant2022universal, li2021statistical, lunt2021measurement, PhysRevB.101.104301, jian2020measurement, li2021conformal, li2021statisticalmech,  liu2022measurement}, which describe the underlying entanglement structure in spacetime. 

Interestingly, the competition between entangling and disentangling effects can arise not only from unitary gates and measurements. 
In fact, phase transitions can also be induced by non-commuting single-qubit measurements and multi-qubit measurements \cite{PhysRevX.11.011030,2020PhRvB.102i4204L, kuno2022emergence,  kuno2023production, roser2023decoding, sang2021entanglement, PhysRevResearch.3.023200, PhysRevResearch.2.023288, lunt2021measurement, kells2023topological, li2021robust, klocke2022topological, 2023arXiv230518559K, 2023arXiv230600058T}, even in the absence of independent unitary gates. 
This scenario is termed as a measurement-only (MO) circuit (MOC) or a projective model in the literature. 
For instance, Lang and B\"{u}chler explored a model called the projective transverse-field Ising (pTFI) model in Ref.~\cite{2020PhRvB.102i4204L} \rev{(see also its generalizations~\cite{kuno2022emergence, kuno2023production, zhu2023structured, kells2023topological, klocke2022topological, roser2023decoding, li2021robust})}. 
In this model, the projective measurement bases consist of the single-qubit $X$ basis and the two-qubit $ZZ$ basis for adjacent sites. 
Each quantum trajectory of the state evolves according to a randomized measurement pattern with these bases, and the entanglement measure, such as the von Neumann entropy, is computed for each trajectory, followed by an average over samples of randomized measurement patterns. 
The averaged entanglement measure reaches an equilibrium (with respect to time steps), and the measurement ratio determines whether the state is in the trivial order or in the \rev{Greenberger-Horne-Zeilinger (GHZ)} order on average.
\rev{It was shown that this $(1+1)$d pTFI model can be mapped to the 2d classical bond percolation model. The criticality extracted from the mutual information was shown to be consistent with this observation.}

From a broader perspective, MOC models may be seen as specialized models of Measurement-Based Quantum Computation (MBQC)~\cite{raussendorf2001one,raussendorf2003measurement,briegel2009measurement,wei2018quantum,wei2021measurement}.
Indeed, Nielsen showed that the universal quantum computation can be achieved by single- and multi-qubit measurements alone~\cite{NIELSEN200396} (see also the review~\cite{wei2021measurement}).
We remark that MOC models utilize measurements with the Pauli or Pauli product basis, and as such, they do not require adaptive determination of parameters for measurement bases unlike in the universal MBQC.
The perspective from MBQC, although not our main focus, will be used at some point in our paper.

In parallel to the development in MIPT, the Higgs phase of gauge theory with matter has gained interest from the viewpoint of the landscape of quantum phases. 
The phase diagram of the gauge-Higgs model contains the confining, deconfining, and Higgs phases~\cite{PhysRevD.19.3682,tupitsyn2010topological}. 
For the model without boundaries, Fradkin and Shenker showed that there is no clear transition between Higgs and confining phases; the two are analytically connected.
If we consider the model with certain types of boundaries, however, the Fradkin-Shenker model was recently shown to experience a phase transition~\cite{verresen2022Higgs}. 
The model hosts degenerate modes on the boundary of the lattice when the model parameters are so that the open Wilson line operator, whose endpoints are attached with charges,
has non-vanishing values in the bulk.
The Higgs phase of the gauge theory 
\rev{has recently been}
interpreted as an SPT phase protected by matter and magnetic symmetries~\cite{verresen2022Higgs}. 
\rev{A salient feature of the model, in particular, in two dimensions is that the supporting mechanism of the boundary degeneracy involves the so-called higher-form symmetry, which is absent in one dimension.} 
\rev{In this work, we address whether such deconfinement-confinement-Higgs phase transitions featured in two dimensions can be observed in an MIPT/MOC setup.}

\rev{Let us review previous studies related to this question. 
Lavasani et al.~\cite{2021NatPh..17..342L} has studied, in a certain parameter limit where no unitary gates are involved, a measurement-only circuit in (1+1)d with measurement bases mimicking the Hamiltonian terms in the 1d cluster SPT Hamiltonian, whose criticality can be mapped to two copies of the 2d bond percolation. Refereces~\cite{klocke2022topological,morral2023detecting} 
also generalize this to
the XZZX model.
More recently, Kuno and Ichinose~\cite{kuno2023production} also studied a generalized setup in (1+1)d but emphasized the `Higgs=SPT' perspective.
Lavasani et al.~\cite{lavasani2021topological} also extended their work to (2+1)d, and studied entanglement transitions involving intrinsic topological orders, i.e., the toric code. 
It was shown that their MOC model can be mapped to a percolation model on the 3d cubic lattice. 
Another intriguing study in Ref.~\cite{kuno2023bulk} found a boundary entanglement transition in the toric code state undergoing a single round of single-qubit projective measurements.}

\rev{
Our work here studies the entanglement transitions in the (2+1)d measurement-only circuit after sufficiently repeated measurement rounds.
The main contributions of the present work are to map out the entire phase diagram in a two-dimensional (probability) parameter space, which involves both bulk and boundary entanglement transitions, and to relate their criticality to that of percolation.
Combining detailed numerical analysis in key limits with an overall scan over the entire parameter space, we map out the phase diagram of the Fradkin-Shenker MOC (FS-MOC) entanglement transitions.}

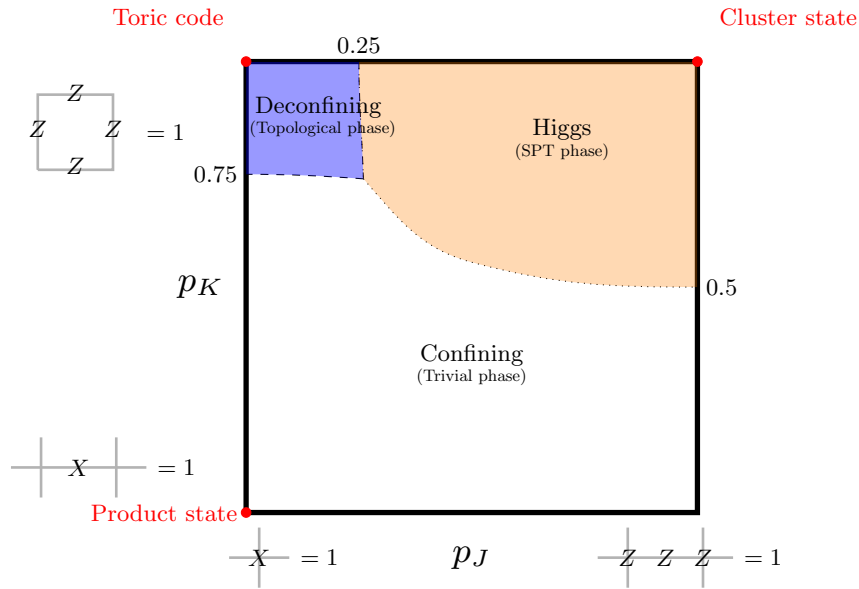
\begin{figure*}

\begin{tikzpicture}[scale=0.6,
  path A/.style 2 args={insert path={(2.5,10) to[out=270, in = 92] #1 (2.6,7.4) #2}},
  path B/.style 2 args={insert path={(2.6,7.4) to[out=311,in=165] #1 (5.0,5.55) #2}},
  path C/.style 2 args={insert path={(5.0,5.55) to[out=-15,in=180] #1 (10,5) #2}},
   path D/.style 2 args={insert path={(10,5) to[out=90,in=270] #1 (10,10) #2}},
   path E/.style 2 args={insert path={(10,10) to[out=180,in=0] #1 (2.5,10) #2}},
   path F/.style 2 args={insert path={(2.6,7.4) to[out=176,in=0] #1 (0,7.5) #2}},
  ]

\draw [-,line width=2.0] (0,0)--(0,10)--(10,10)--(10,0)--(0,0);
\node[scale=1.5] at (5,-1.0) {$p_J$};
\node[scale=1.5] at (-1.0,5) {$p_K$} ;

    \draw[path A={coordinate[pos=0] (A1)
     coordinate[pos=1] (A2)}{},dotted];
    \draw[path A={coordinate[pos=0] (A1)
     coordinate[pos=1] (A2)}{},dashed];
     
    \draw[path B={coordinate[pos=0] (B1) 
        coordinate[pos=1] (B2)}{},dotted];
    \draw[path C={coordinate[pos=0] (C1)
    coordinate[pos=1] (C2)}{},dotted];
    
    \draw[path D={coordinate[pos=0] (D1)
    coordinate[pos=1] (D2)}{},dashed];
    \draw[path E={coordinate[pos=0] (E1)
    coordinate[pos=1] (E2)}{},dashed];
    \draw[path F={coordinate[pos=0] (F1)
    coordinate[pos=1] (F2)}{},dashed];
    
    \begin{scope}
     \clip[path A={}{-- (0,0) -- (10,0) -- (10,10) -- cycle}];
     \clip[path B={}{-- (10,0)-- (10,10) -- (1.0,10) -- cycle}];
     \clip[path C={}{-- (10,10) -- (1,10) -- (1,5) -- cycle}];
     \fill[orange,opacity=0.3] (0,0) -- (10,0) -- (10,10) -- (0,10) -- cycle;
    \end{scope}

    \begin{scope}
    \clip[path A = {}{ -- (0,0.7) -- (0,10) -- cycle}];
    \clip[path F ={}{--(0,10)--(3,10)-- cycle}];
    \fill[blue,opacity=0.4] (0,0) -- (10,0) -- (10,10) -- (0,10) -- cycle;
    \end{scope}

    \node at (1.6,9) {Deconfining};
     \node[scale =0.7] at (1.7,8.5) {(Topological phase)};

     \node at (7,8.5) {Higgs};
    \node[scale =0.7] at (7,8.0) {(SPT phase)};
     
    \node at (5,3.5) {Confining};
    \node[scale =0.7] at (5,3.0) {(Trivial phase)};
    
    \filldraw[red] (10,10) circle (3.0pt);
	\node[red,right,yshift=0] at  (10.3,11) {Cluster state};

     \filldraw[red] (0,10) circle (3.0pt);
	\node[red,left,yshift=0] at  (-0.3,11) {Toric code};

 \filldraw[red] (0,0) circle (3.0pt);
	\node[red,left,yshift=0] at  (0.0,0) {Product state};
 
    \node[right,yshift=0] at  (10,5) {0.5};

     \node[left,yshift=0] at  (0,7.5) {0.75};

     \node[above,yshift=0] at  (2.5,10) {0.25};

     \node at (10,-1){
    \raisebox{-10pt}{\begin{tikzpicture}[scale=1.0]
        \draw[-,black!30,line width=1.0] (-0.4,0.0) -- (1.4,0.0);
        \draw[-,black!30,line width=1.0] (0.0,-0.4) -- (0.0,0.4);
        \draw[-,black!30,line width=1.0] (1.0,-0.4) -- (1.0,0.4);
        \node[scale=1.0] at (1.0,0.0) {$Z$};
        \node[scale=1.0] at (0.5,0.0) {$Z$};
        \node[scale=1.0] at (0.0,0.0) {$Z$};
        \node[scale=1.0] at (1.8,0) {$=1$};
        \end{tikzpicture} }
     } ;
    \node at (1,-1){
     \raisebox{-10pt}{\begin{tikzpicture}[scale=1.0]
        \draw[-,black!30,line width=1.0] (-0.4,0.0) -- (0.4,0.0);
        \draw[-,black!30,line width=1.0] (0.0,-0.4) -- (0.0,0.4);
        \node[scale=1.0] at (0.0,0.0) {$X$};
        \node[scale=1.0] at (0.8,0) {$=1$};
        \end{tikzpicture} }
     } ;

     \node at (-3,1){
     \raisebox{-10pt}{\begin{tikzpicture}[scale=1.0]
        \draw[-,black!30,line width=1.0] (-0.4,0.0) -- (1.4,0.0);
        \draw[-,black!30,line width=1.0] (0.0,-0.4) -- (0.0,0.4);
        \draw[-,black!30,line width=1.0] (1.0,-0.4) -- (1.0,0.4);
        \node[scale=1.0] at (0.5,0.0) {$X$};
        \node[scale=1.0] at (1.8,0) {$=1$};
        \end{tikzpicture} }
     } ;

      \node at (-3,8.5){
             \raisebox{-10pt}{\begin{tikzpicture}[scale=1.0]
                \draw[-,black!30,line width=1.0] (-0.5,-0.5) -- (0.5,-0.5) -- (0.5,0.5) -- (-0.5,0.5) -- (-0.5,-0.5);
        \node at (-0.5,0.05) {$Z$};
        \node at (0.5,0.05) {$Z$};
        \node at (0.0,0.53) {$Z$};
        \node at (0.0,-0.45) {$Z$};
        \node[scale=1.0] at (1.2,0) {$=1$};
                \end{tikzpicture} }
     } ;

\end{tikzpicture}
\caption{A schematic structure of the phase diagram in the projective gauge-Higgs MOC model. 
\rev{
{\bf Deconfining phase:} Transitions from the deconfining phase to others are probed by the topological entanglement entropy. 
We claim that the transition point $(p_J,p_K)=(0.25,1)$ (see Fig.~\ref{fig:cut_i} and Fig.~\ref{fig:cut_i_half}) is described by the bond percolation on the 3d cubic lattice.
We also claim that the transition point $(p_J,p_K)=(0,0.75)$ (see Fig.~\ref{fig:cut_ii}) is described by the so-called homological percolation on the 3d cubic lattice.
{\bf Higgs phase:} Transitions from the Higgs phase to others are detected by the boundary mutual information. 
We claim that the transition point $(p_J,p_K)=(1,0.5)$ (see Fig.~\ref{fig:cut_v}) is described by the bond percolation on the 2d square lattice.
It is not conclusive in this study whether or not the Higgs phase meets the deconfining phase or the self-duality line $p_K = 1-p_J$.
}
}
\label{fig:big-picture}
\end{figure*}

\rev{
More concretely, we study an MOC whose measurement bases mimic the Hamiltonian terms in the $(2+1)$d Fradkin and Shenker's gauge-Higgs model on a 2d lattice with a boundary.
We obtain a phase diagram, parameterized by two measurement ratios, that contains three phases as in the Hamiltonian model---the topological order (deconfining phase), the SPT order (Higgs phase), and the trivial order  (confining phase); see Fig.~\ref{fig:big-picture} for the summary.
As a diagnostic of the topological order in the MOC, we compute the topological entanglement entropy~\cite{kitaev2006topological, satzinger2021realizing} in the bulk and show that there is a region with constant topological entanglement and a clear phase transition separating this region from the rest.
As a diagnostic of the SPT order, on the other hand, we will essentially diagnose a GHZ state induced at the boundary of the lattice. 
To do so, we compute the mutual information between a pair of boundary regions; our idea traces back to the original study by Lang and B\"{u}chler, who studied the phase transition in $(1+1)$d pTFI model using the mutual information~\cite{2020PhRvB.102i4204L}.
Our numerical result shows a clear phase boundary separating a parameter region corresponding to the Higgs phase from the rest, which is in agreement with the recent proposition of `Higgs=SPT'~\cite{verresen2022Higgs} in the Hamiltonian picture.}
\rev{
We stress that existence of such a parameter region is a non-trivial result in this work.
}

\rev{
We further provide more detailed understandings on some key limits in the phase diagram.
Namely, criticality in certain limits extracted from our numerics reproduces some of the results in previous works.
Below we summarize them:
\begin{itemize}
\item In a limit (an SPT-trivial transition), the boundary system can be viewed as decoupled from the bulk through a duality map. We show that the criticality is consistent with that of the $(1+1)$ pTFI model by Lang and B\"uchler~\cite{2020PhRvB.102i4204L}, which is underpinned by the 2d bond percolation on the 2d square lattice. 
\item In another limit (a deconfining-Higgs transition), the Fradkin-Shenker MOC model reduces to the $(2+1)$d pTFI model~\cite{2020PhRvB.102i4204L}, where we find that the criticality exhibited is consistent with that of the bond percolation on the 3d cubic lattice.
\item In yet another limit (a deconfining-confining transition), the Fradkin-Shenker MOC model reduces to a pure gauge MOC model. We numerically find that the critical threshold is consistent with that of the so-called homological percolation on the 3d cubic lattice.
Theoretically, we relate the pure gauge MOC and the homological percolation using MBQC in Appendix~\ref{sec:RBH};
namely, we explicitly give an MBQC realization of the (2+1)d pure-gauge MOC model as sequential, randomized single-qubit measurements on the 3d cluster state by Raussendorf, Bravyi, and Harrington~\cite{raussendorf2005long}.
It is then geometrically evident that the transition is driven by the surface percolation in the 3d cubic lattice.
\end{itemize}
Regarding the last point, we comment that the full Fradkin-Shenker MOC model in the main text can be derived by generalizing the discussion in Appendix~\ref{sec:RBH} to a 3d state constructed by one of us in Ref.~\cite{Okuda-Parayil-Mana-Sukeno}, where various but different theoretical aspects in single-qubit projective measurements on the 3d entangled state were discussed. 
}

\rev{Apart from the main entanglement analysis, we will also show that our MO circuit hosts a ``mixed phase''---an analogue of the water-vapor phase (it should be distinguished from the so-called mixed-state quantum phase)---which is characterized by alternating appearance of the cluster state and the product state  in the bulk (at one specific corner of the phase diagram), depending on how we end our circuit. }
This will be detected by a non-local bulk quantity.
Remarkably, this observable exhibits continuity between the Higgs limit and the confining limit, whereas it also shows a clear separation from the deconfining phase---a picture consistent with the original proposition by Fradkin and Shenker.
\rev{In Appendix~\ref{sec:XZ-randomized}, we discuss an alternative MOC that has a random temporal structure.}

This paper is organized as follows.
\rev{In Section~\ref{section:pHG-MOC}, we describe our Fradkin-Shenker MOC problem.
We first define our setup and then discuss the diagnostics of quantum orders in our MOC. 
In Section~\ref{sec:results}, we provide our numerical results and present our interpretation in terms of percolation. }
Section~\ref{sec:conc-diss} is devoted to conclusions and discussion.

\section{Projective-gauge-Higgs MOC}
\label{section:pHG-MOC}

\rev{
\subsection{Review: Hamiltonian formulation of the Fradkin-Shenker model}
}

\rev{Let us first briefly review the Fradkin-Shenker model in the quantum Hamiltonian setup and fix some notations. 
We will introduce our Fradkin-Shenker MOC model in the next subsection.
}

\begin{figure*}
\begin{center}
\includegraphics[width=1.0\linewidth]{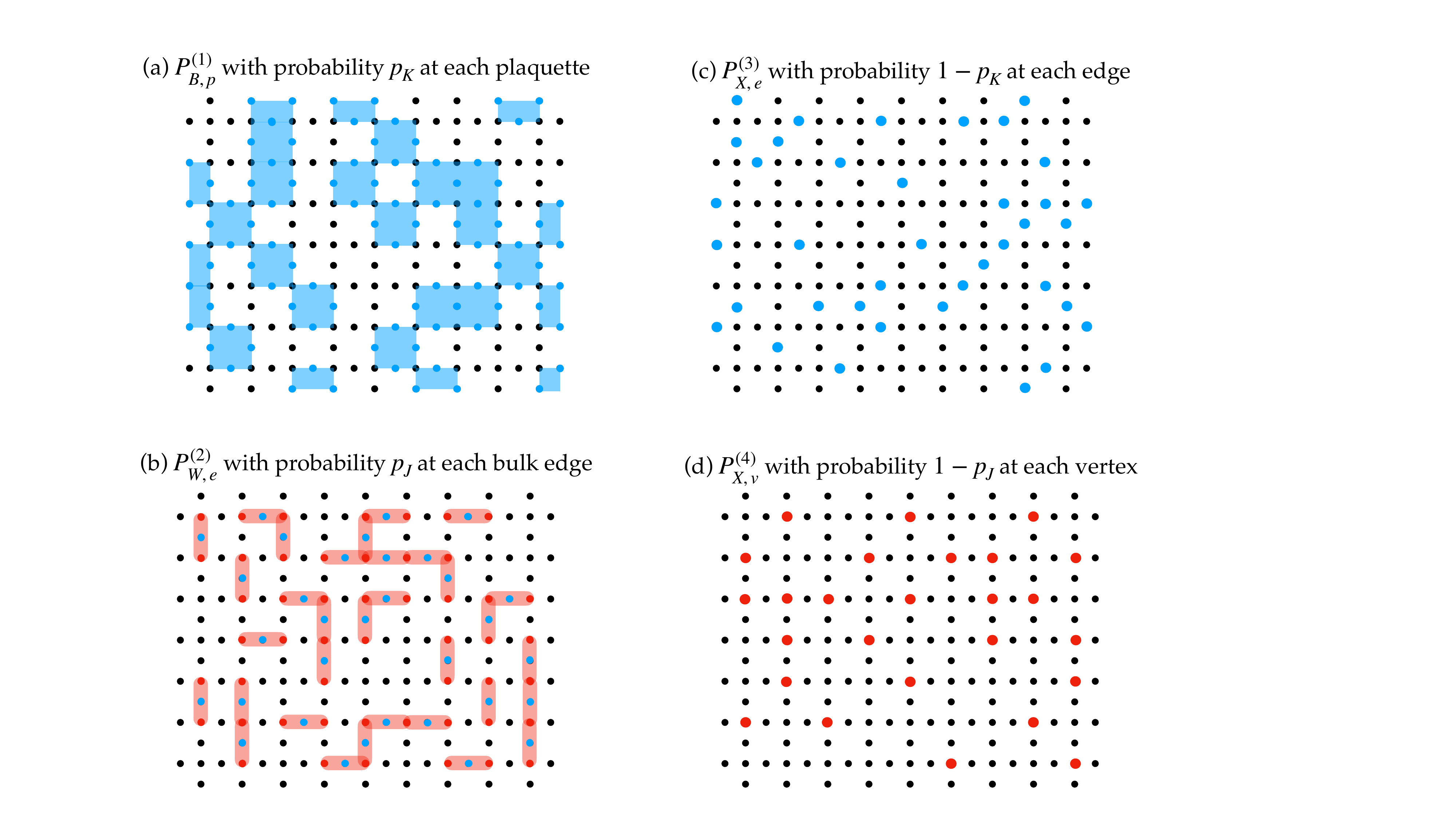}
\caption{Setup of our projective gauge-Higgs MOC model. (a) Step (1-1) of the algorithm. Each plaquette is activated with probability $p_K$ (blue) and then the operator $B_p$ is measured. At the rough boundary, the $B_p$ operator is three bodies on sides and two bodies at the four corners. (b) Step (1-2) of the algorithm. Each edge is activated with probability $p_J$ (red) and then the operator $W_e$ is measured. Note that the rough boundary is not involved in this measurement. (c) Step (2-1) of the algorithm. Each edge is activated with probability $1-p_K$ (blue) and then the operator $X_e$ is measured. (d) Step (2-2) of the algorithm. Each vertex is activated with probability $1-p_J$ (red) and then the operator $X_v$ is measured.}
\label{fig:FS-MIPT-setup}
\end{center}
\end{figure*}

Let $X$ and $Z$ be the Pauli $X$  and $Z$ operators, respectively. 
We denote their eigenvectors as
\begin{align}
&Z|0 \rangle = | 0\rangle, \quad Z|1\rangle = - |1 \rangle , \\
&X|+ \rangle = | +\rangle, \quad X|-\rangle = - |- \rangle , \\
&|+\rangle = \frac{|0 \rangle + |1\rangle}{\sqrt{2}} , \quad 
|-\rangle = \frac{|0 \rangle - |1\rangle}{\sqrt{2}} .
\end{align}
The Hilbert space is defined as the tensor product of qubits placed on vertices $V$ and edges $E$ of the 2d square lattice; this is also called the Lieb lattice.
We consider primarily the Lieb lattice with the so-called rough boundary so that the boundary of the lattice consists of edges perpendicular to it.  (We have also considered the smooth boundary condition but remark that the rough boundary allows us to better detect the transition via mutual information. The choice of the rough boundary was motivated by the setup in the work `Higgs=SPT' by Verresen and collaborators~\cite{verresen2022Higgs}.) 
We have three types of cells: plaquettes $P$, edges $E$, and vertices $V$. 
When clarification is needed, we use subscripts $B$ and $\partial$ to indicate the set of cells is restricted to the bulk and the boundary, respectively.
For instance, $E_B$ denotes the set of all edges in the bulk, $E_\partial$ denotes the set of all edges on the boundary, $E_{B\partial}$ denotes the set of all of those in the bulk or on the boundary, and so on. 
See Fig.~\ref{fig:FS-MIPT-setup} for illustration of our setup.

\rev{We introduce the following notations:}\footnote{%
For example, an edge pointing in the $x$ direction can be expressed as $e=\{ (t,y) |  x\leq t \leq x+1\}$, and its boundary includes vertices at points $\{(x,y)\}$ and $\{(x+1,y)\}$, which are subsets of $e$. } 
\begin{align}
W_e &= Z_e \prod_{v \subset e} Z_v
=\raisebox{-10pt}{\begin{tikzpicture}
\draw[-,black!30,line width=1.0] (-0.4,0.0) -- (1.4,0.0);
\draw[-,black!30,line width=1.0] (0.0,-0.4) -- (0.0,0.4);
\draw[-,black!30,line width=1.0] (1.0,-0.4) -- (1.0,0.4);
\node at (1.0,0.0) {$Z$};
\node at (0.5,0.0) {$Z$};
\node at (0.0,0.0) {$Z$};
\end{tikzpicture} }
\quad \forall e \in E_B, \\
B_p &= \prod_{e \subset p} Z_e 
=
\raisebox{-20pt}{\begin{tikzpicture}
\draw[-,black!30,line width=1.0] (-0.5,-0.5) -- (0.5,-0.5) -- (0.5,0.5) -- (-0.5,0.5) -- (-0.5,-0.5);
\node at (-0.5,0.05) {$Z$};
\node at (0.5,0.05) {$Z$};
\node at (0.0,0.53) {$Z$};
\node at (0.0,-0.45) {$Z$};
\end{tikzpicture} }
\quad \forall p \in P_{B} 
. 
\end{align}
The term $W_e$ describes a coupling among the gauge degrees of freedom on an edge and the two matter degrees of freedom on the vertices adjacent to it. 
Note that we do not introduce such a term 
for edges on the rough boundary, as each of them only contains a single vertex. 
On the other hand, the term $B_p$ is the plaquette term, which describes the magnetic interaction in the gauge theory.
The plaquette term in the bulk is the product over the four gauge degrees of freedom around a plaquette.
On the rough boundary, we define it as the product over three gauge degrees of freedom. 
At the four corners of the lattice, we define it as the product over the two edges that sit at the corner.
We also write the Gauss law divergence operator as
\begin{align}
G_v = X_v \prod_{e \supset v} X_e = 
\raisebox{-18pt}{\begin{tikzpicture}
\draw[-,black!30,line width=1.0] (0.0,0.0) -- (1.0,0.0);
\draw[-,black!30,line width=1.0] (0.5,-0.5) -- (0.5,0.5);
\node at (0.0,0.0) {$X$};
\node at (0.5,0.0) {$X$};
\node at (1.0,0.0) {$X$};
\node at (0.5,0.5) {$X$};
\node at (0.5,-0.5) {$X$};
\end{tikzpicture} }
\, .
\end{align}

The Hamiltonian of the (2+1)d Fradkin-Shenker model is given by~\cite{PhysRevD.19.3682}
\begin{align} \label{eq:FS-Hamiltonian}
H=& 
- J \sum_{e \in E_{B} } W_e
- h_1 \sum_{v \in V } X_v \nonumber \\ 
&- K \sum_{p \in P_{B\partial}} B_p
- h_2 \sum_{e \in E_{B\partial}} X_e  .
\end{align}
The model on a periodic lattice was shown to possess the deconfining phase at the large $(h_1,K)$ limit, which is said to have a long-range topological order~\cite{wen1990topological, chen2010local}, and other two (confining and Higgs) phases that are smoothly connected. The model corresponds to a 2d quantum model, which is a toric code in external fields obtained by fixing the state to the so-called unitary gauge. 

\rev{The phase diagram of the corresponding 3d classical model  was studied by Jongeward et al.~\cite{PhysRevD.21.3360} and the quantum model   was later investigated with different methods~\cite{2009PhRvB..79c3109V,tupitsyn2010topological, 2012PhRvB..85s5104W}. 
Additionally, there was a study from the perspective of the wave function deforming the toric code~\cite{2019PhRvL.122q6401Z}.
We mention that the corresponding 3d classical statistical model is also discussed from the perspective of percolation~\cite{PhysRevD.72.054509,linsel2024percolation}.
}

\rev{
When the FS model is put on a lattice with the rough boundary, Verresen et al.~\cite{verresen2022Higgs} recently showed that the model exhibits a phase transition between the Higgs (in the region with large $(J,K)$) and the confining phases (in the region with large $(h_1,h_2)$), which is signaled by a spontaneous symmetry breaking detected by a local order parameter placed at the boundary. 
As we will review later, the disorder parameter of the Higgs phase
is a non-local, gauge-invariant order parameter in the bulk, and it is regarded as a string order parameter for an SPT phase~\footnote{%
In the Higgs phase, the magnetic 1-form symmetry is preserved, and the open Wilson line operator, serving as a disorder operator, has a non-vanishing VEV~\cite{verresen2022Higgs}. 
}.
From the perspective of the lattice gauge theory, it is an open Wilson line operator, which is a product of $W_e$ over several edges. 
On the other hand, from the modern condensed matter or quantum-information theory perspective, it is a product of stabilizers of the so-called cluster state. 
Indeed, the Higgs=SPT limit describes the 2d cluster state on the Lieb lattice. 
}

\rev{
\subsection{Our setup: Fradkin-Shenker measurement-only circuit}
}

Now, we introduce a stabilizer circuit implemented by a set of local measurements that mimic the terms in the above Hamiltonian.
We do so by writing the projectors:
\begin{align}
P^{(1)}_{B,p} &= \frac{1+ B_p}{2} \quad \forall p \in P_{B\partial}, \\
P^{(2)}_{W,e} &= \frac{1+ W_e}{2}  \quad \forall e \in E_{B}, \\
P^{(3)}_{X,e} &= \frac{1+ X_e}{2} \quad \forall e \in E_{B\partial}, \\
P^{(4)}_{X,v} &= \frac{1+ X_v}{2} \quad \forall v \in V ,
\end{align}
where the measurement outcome was fixed to $+1$ by setting the coefficients of the operators to be $+1$, for the reason we explain below. 
Note that all the projectors commute with the Gauss law generator: $[P^{(\bullet)}_\bullet, G_v]=0$ for all $v \in V$, and therefore, the Gauss law is preserved in our simulation (provided we start with a state respects it). 

Indeed, we initialize the state on the 2d Lieb lattice as 
\begin{align}
|\psi_{\text{ini}}\rangle = |+\rangle^{\otimes E} |+\rangle^{\otimes V} ,
\end{align}
which satisfies the Gauss law constraint $G_v = 1$.
In other words, we take the initial stabilizer generators as
\begin{align}
S[0] = \big\langle X_{v} , X_{e} \big| v \in V , e \in E \big\rangle . 
\end{align}
Now the measurement pattern for gauge-Higgs MOC is as follows (see Fig.~\ref{fig:FS-MIPT-setup} for illustration):
\begin{enumerate}
\item[] \hspace{-20pt} \underline{ {\bf Algorithm (projective-gauge-Higgs MOC)} }
\item[(0)] Initialize a 2d state in the product state $|+\rangle^{\otimes V} |+\rangle^{\otimes E}$.
\item[(1-1)] Associated with each face, we perform the measurement in $P^{(1)}_{B,p}$ with probability $p_K$.
\item[(1-2)] Associated with each edge, we perform the measurement in $P^{(2)}_{W,e}$ with probability $p_J$.
\item[(2-1)] Associated with each edge, we perform the measurement in $P^{(3)}_{X,e}$ with probability $1-p_K$.
\item[(2-2)] Associated with each vertex, we perform the measurement in $P^{(4)}_{X,v}$ with probability $1-p_J$.
\item[(3)] Repeat (1)-(2) for $N_t$ rounds. 
\item[(4)] Compute entanglement measures on the resulting state.
\item[(5)] Repeat (0)-(4) for $N_s$ ($N_s$: the number of samples) times. Compute the average of the entanglement measures over the samples.
\end{enumerate}

The MOC starts with a Pauli stabilizer state and evolves by measurement of Pauli operators, so it is described by the so-called stabilizer circuit (see e.g. Ref.~\cite{aaronson2004improved}), and it can be efficiently simulated on classical computers.
The state after the each measurement-based evolution (1)-(3) is again a stabilizer state.
The precise stabilizers of the state depends on both which measurement basis was chosen at each randomized measurement and which measurement outcome we obtain ($+1$ or $-1$; with probability $\frac{1}{2}$ each when not deterministic). 
The latter dependency of $\pm$ outcomes, however, does not affect the value of entanglement measures (such as the entanglement entropy) since the difference in measurement outcomes can be accounted for by applying local Pauli operators on the stabilizer state with all $+1$ outcomes.
This is why we can restrict our projectors to those with $+1$ measurement outcomes in our simulations.

\subsection{Phases and their diagnostics}

\rev{Let us explain how we characterize phases in our FS-MOC problem.}

First,  when $p_K \simeq 1$ and $p_J \simeq 1$ so that measurements with $P^{(1)}_{B,p}$ and $P^{(2)}_{W,e}$ dominate, the simulated state would flow to an SPT state. 
This is a state whose stabilizers are $W_e$ and $G_v$. 
(Note that $W_e=+1$ for all $e\in E_B$ implies $B_p =+1$.)
Up to the Hadamard transformation, this is the same as the cluster state on the Lieb lattice, which possesses an SPT order protected by $\mathbb{Z}_2[1] \times \mathbb{Z}_2[0]$, see \cite{ yoshida2016topological,verresen2022Higgs,sukeno2022measurementbased, li2023measuring} for examples. This SPT phase was recently interpreted as a Higgs phase by Verresen et al.~\cite{verresen2022Higgs}.
\rev{As we already mentioned, the confining and the Higgs phases are smoothly connected in the quantum Hamiltonian formulation without boundaries, and there is no clear phase boundary between them as far as bulk observables are concerned. On the other hand, the Higgs=SPT phase, which is separated by a phase boundary from the rest of phases, can be diagnosed by a robust degeneracy at the rough boundary in the quantum Hamiltonian spectrum~\cite{verresen2022Higgs}. 
Motivated by the quantum Hamiltonian perspective, we detect the Higgs=SPT phase in our projective model with what we call bondary mutual information.}

Secondly, when $p_K \simeq 1$ and $p_J \simeq 0$, the simulated state would keep being measured in the bases $P^{(1)}_{B,p}$ and $P^{(4)}_{X,v}$, and the state flows to a product of the trivial state on vertices and the toric code on edges. 
This state exhibits a non-trivial topological order.
\rev{In the quantum Hamiltonian formulation, the deconfining phase is robustly characterized by its underlying topological data.
Hence, we employ the topological entanglement entropy to detect an analogous phase in the measurement-only setup.}

Thirdly, when $p_K \simeq 0$ and $p_J \simeq 0$ so that measurements with $P^{(3)}_{X,e}$ and $P^{(4)}_{X,v}$ are performed often, the state would flow to a trivial product state. 

Finally, at the limit $p_K \simeq 0$ and $p_J \simeq 1$, the boundary degrees of freedom are frozen to $X_e =1$, while the decoupled bulk alternates between the cluster state at Steps~(1-1)~\&~(1-2) and the product state at Steps~(2-1)~\&~(2-2). 
This `mixed-phase' regime corresponds to the highly frustrated ground state in the Hamiltonian~\eqref{eq:FS-Hamiltonian} with the $W_e$ term and the $X_e$ term being both present, the latter of which explicitly breaks the magnetic 1-form symmetry $\mathbb{Z}_2[1]$.
\rev{The mixed phase in our model appears artificially as we separate non-commuting terms in Hamitonian into different measurement rounds.}
\rev{The alternating bulk states will be exhibited by another (step-sensitive) operator (the open Wilson line operator), which we introduce below as well.}

Later, we will numerically show that the mixed-phase behavior characterized by alternating bulk states does not show up in 
\rev{the topological entanglement entropy and the boundary mutual information,}
i.e., the essential features of them do not change significantly whether we stop after full cycles (after Steps (1-1), (1-2), (2-1), and (2-2)) or stop at the middle (after Steps (1-1) and (1-2)). 
We illustrate our schematic phase diagram characterized by the robust entanglement measures in Fig.~\ref{fig:big-picture}.

\subsubsection{Topological entanglement entropy}

\begin{figure*}
\includegraphics[width=0.6\linewidth]{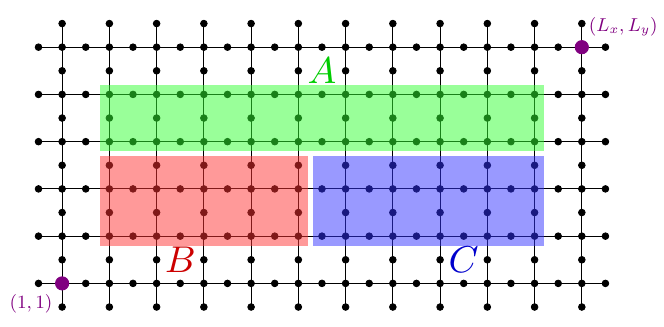}
\caption{The set up for the topological entanglement entropy $S_\text{topo}$. We label vertices by integer coordinates. The above figure is for $(L_x,L_y)=(12,6)$. The region $A$ is bounded by the square whose southwest corner is at $(2,L_y\div 2 +1 )$ and northeast corner at $(L_x-1,L_y-1)$. The region $B$ is bounded by a square whose southwest corner is at $(2,2)$, southeast corner at $(L_x\div 2 , 2)$, and appends gauge qubits below the region $A$. The region $C$ is bounded by a square whose southeast corner is at $(L_x-1, 2)$ and fills the quadrant formed in the southeast of $A$ and $B$. }
\label{fig:TEE_setup}
\end{figure*}

To detect the topological order, we compute the topological entanglement entropy, in particular, the setup by Kitaev and Preskill~\cite{kitaev2006topological}, which is a tripartite entanglement measure among three regions, $A$, $B$, and $C$.
Let $S_R$ be the entanglement entropy for the region $R$ ($R \in \{A,B,C\}$). 
Then the topological entanglement entropy (TEE) is given by 
\begin{align}
S_{\rm topo} = S_A + S_B +S_C - S_{AB} - S_{BC} - S_{AC} + S_{ABC} . 
\end{align}
We illustrate our setup of regions $A$, $B$, and $C$ in Fig.~\ref{fig:TEE_setup}.
In the toric code limit ($p_J \simeq 0$, $p_K \simeq 1$), each entanglement entropy term contains the area law term as well as the constant $\gamma$ that characterizes the topological order, but in $S_{\rm topo}$ the area terms cancel one another so it reduces to the topological constant $\gamma = -1$ (in base 2 of logarithm).
As explained in Ref.~\cite{hamma2005bipartite,nahum2017quantum} and in Appendix~\ref{sec:EE-stabilizer}, calculation of the entanglement entropy can be done efficiently for stabilizer circuits.

\subsubsection{Boundary mutual information}

Now we explain that the mutual information between two regions, each of which are at the vicinity of rough boundaries, can be used to detect the Higgs=SPT phase. 
To derive it, we begin by reviewing the idea by Verresen et al.~\cite{verresen2022Higgs}.

The Higgs phase is characterized by 
the non-vanisning open Wilson line operator
\begin{align}
\left\langle 
Z_v \cdot \prod_{e \in \Gamma} Z_e \cdot Z_{v'}
\right\rangle \neq 0 ,
\end{align}
with an open line $\Gamma \subset E_B$ and the endpoints of $\Gamma$ being $v$ and $v'$. 
In the context of SPT phases, this is the string order parameter whose endpoints are charged under the matter $\mathbb{Z}_2$ symmetry generated by $P = \prod_{v\in V} X_v$.
In the same phase, we also have the magnetic 1-form symmetry generated by the operators on a line or a loop $\gamma \subset E_{B\partial}$: $W_\gamma = \prod_{e \in \gamma} Z_e$, which is a product of $B_p$'s. 
Note that $\gamma$ is either a closed loop in the bulk or a line that terminates on the rough boundary. 

Let us consider a particular open Wilson line operator
whose endpoints are close to the rough boundary: 
\begin{center}
\begin{tikzpicture}
\node at (0,0){
    \begin{tikzpicture}
	\draw[-,dotted,black!80] (-.7,.5-2) -- (4+.7,.5-2);
	\foreach \x in {2,3}{
		\draw[-] (-.7,0-\x) -- (4+.7,0-\x);
	}
	\foreach \x in {0,1,2,3,4}{
		\draw[-] (\x,.7-2-0.2) -- (\x,-3.7);
	};
	\foreach \x in {0,1,2,3,4}{
		\filldraw (\x,-1.5) circle (2pt);
		\filldraw (\x,-2) circle (2pt);
		\filldraw (\x,-2.5) circle (2pt);
		\filldraw (\x,-3) circle (2pt);
		\filldraw (\x,-3.5) circle (2pt);
	};
	\foreach \x in {0,1,2,3,4,5}{
		\filldraw (\x-0.5,-2) circle (2pt);
		\filldraw (\x-0.5,-3) circle (2pt);
	};
	\draw[blue,snake it] (0,-2-1) -- (4,-2-1);
	\foreach \x in {0,1,2,3}{
		\filldraw[blue] (0.5+\x,-2-1) circle (2.1pt);
		\node[blue,below,yshift=0] at (0.5+\x,-2-1) {$Z$};
	};
 	\draw[blue,snake it] (0.0,-2) -- (0.0,-3);
    \draw[blue,snake it] (4.0,-2) -- (4.0,-3);
    
    \filldraw[blue] (0.0,-2.5) circle (2.1pt);
	\node[blue,left,yshift=0] at  (0.0,-2.5) {$Z$};

     \filldraw[blue] (4.0,-2.5) circle (2.1pt);
	\node[blue,right,yshift=0] at  (4.0,-2.5) {$Z$};
 
	\filldraw[color=red] (0,-2) circle (2.1pt) node[above left] {$Z$};
	\filldraw[color=red] (4,-2) circle (2.1pt) node[above right] {$Z$};
	\end{tikzpicture}
};
\end{tikzpicture} 
\end{center}
Following Ref.~\cite{verresen2022Higgs}, we use the magnetic symmetry (both in the bulk and along the boundary) in the SPT phase to rewrite the above operator. 
Then, we get the following operator that has non-vanishing vacuum expectation value:
\begin{center}
\begin{tikzpicture}
\node at (0,0){
    \begin{tikzpicture}
	\draw[-,dotted,black!80] (-.7,.5-2) -- (4+.7,.5-2);
	\foreach \x in {2,3}{
		\draw[-] (-.7,0-\x) -- (4+.7,0-\x);
	}
	\foreach \x in {0,1,2,3,4}{
		\draw[-] (\x,.7-2-0.2) -- (\x,-3.7);
	};
	\foreach \x in {0,1,2,3,4}{
		\filldraw (\x,-1.5) circle (2pt);
		\filldraw (\x,-2) circle (2pt);
		\filldraw (\x,-2.5) circle (2pt);
		\filldraw (\x,-3) circle (2pt);
		\filldraw (\x,-3.5) circle (2pt);
	};
	\foreach \x in {0,1,2,3,4,5}{
		\filldraw (\x-0.5,-2) circle (2pt);
		\filldraw (\x-0.5,-3) circle (2pt);
	};

 	\draw[blue,snake it] (0.0,-2) -- (0.0,-1.5);
    \draw[blue,snake it] (4.0,-2) -- (4.0,-1.5);
    
    \filldraw[blue] (0.0,-1.5) circle (2.1pt);
	\node[blue,left,yshift=5] at  (0.0,-1.5) {$Z$};
    \node[black,above,yshift=0] at  (0.0,-1.0) {region $A$};

     \filldraw[blue] (4.0,-1.5) circle (2.1pt);
	\node[blue,right,yshift=5] at  (4.0,-1.5) {$Z$};
 \node[black,above,yshift=0] at  (4.0,-1.0) {region $B$};

	\filldraw[color=red] (0,-2) circle (2.1pt) node[above left] {$Z$};
	\filldraw[color=red] (4,-2) circle (2.1pt) node[above right] {$Z$};
	\end{tikzpicture}
};
\end{tikzpicture} 
\end{center}
 We denote it as $Z(A,B)=Z(A)Z(B)$, where $Z(A) = Z_{v(A)}Z_{e(A)}$ for the small region $A=\{v(A),e(A)\}$ and a similar definition for $B$. 
Here, $v(A)$ and $e(A)$ are the vertex and the edge that constitute the region $A$, respectively. 

We also observe that our initial state and the measurement bases are symmetric under both the local Gauss law generator and the global $P$ symmetry generator. 
Hence, we get a boundary global symmetry~\cite{verresen2022Higgs}
\begin{align}
1 = P \times \prod_{v \in V} G_v = \prod_{e \in E_\partial} X_e =: P_\partial , 
\end{align}
which can be depicted below.
\begin{center}
\begin{tikzpicture}
\node at (0,0){
    \begin{tikzpicture}
	\draw[-,dotted,black!80] (-.7,.5-2) -- (4+.7,.5-2);
	\foreach \x in {2,3}{
		\draw[-] (-.7,0-\x) -- (4+.7,0-\x);
	}
	\foreach \x in {0,1,2,3,4}{
		\draw[-] (\x,.7-2-0.2) -- (\x,-3.7);
	};
	\foreach \x in {0,1,2,3,4}{
		\filldraw (\x,-1.5) circle (2pt);
		\filldraw (\x,-2) circle (2pt);
		\filldraw (\x,-2.5) circle (2pt);
		\filldraw (\x,-3) circle (2pt);
		\filldraw (\x,-3.5) circle (2pt);
	};
	\foreach \x in {0,1,2,3,4,5}{
		\filldraw (\x-0.5,-2) circle (2pt);
		\filldraw (\x-0.5,-3) circle (2pt);
	};

   \foreach \x in {0,1,2,3,4}{
     \filldraw[blue] (0.0+\x,-1.5) circle (2.1pt);
	\node[blue,above,yshift=3] at  (0.0+\x,-1.5) {$X$};
   }

 	\draw[blue,snake it] (-0.5,-1.5) -- (4.5,-1.5);

	\end{tikzpicture}
};
\end{tikzpicture} 
\end{center}
One can also directly show that $P_\partial$ is a symmetry of all the measurement bases.

Each local product $Z(A)$ or $Z(B)$ is individually charged under the matter's symmetry generator $P_\partial$ --- a hallmark of the symmetry-breaking boundary degeneracy in the Higgs=SPT phase.

\begin{figure}
    \includegraphics[width=1.0\linewidth]{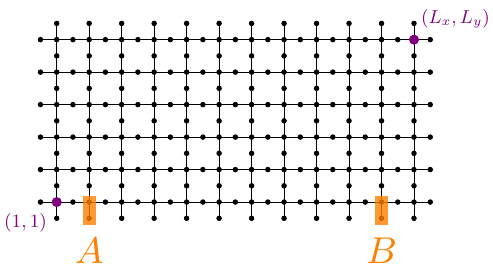}
    \caption{Our setup for the boundary mutual information $I(A:B)$.}
    \label{fig:BMI}
\end{figure}

The discussion above indicates that there are Bell-like pair creations on the boundary.
Indeed, the situation is comparable to the 1d GHZ state whose stabilizers are given by $Z_iZ_j$ ($i<j$) and $\prod_{i} X_i$, except our effective 
\rev{$Z$ operator}
includes a $Z$ at the rough boundary and the $Z$ next to it. 
On the other hand, in the study by Lang and B\"{u}chler, the mutual information was used to detect the GHZ long-range order in the 1d pTFI model.
This motivates us to use the mutual information between the two separate regions close to the boundary, written as $I(A:B)$, as the diagnostic of the SPT phase. 
It is defined as 
\begin{align}
I(A:B) =  S_A  +  S_B  - S_{A\cup B},
\end{align}
and it becomes $I(A:B)\simeq 1$ in the Higgs=SPT-phase limit $(p_J,p_K)\simeq (1,1)$.
We call this diagnostic the \emph{boundary mutual information (BMI)}; see Fig.~\ref{fig:BMI} for the illustration of the setup. 

\subsubsection{Open Wilson line operator in the bulk}

\rev{In the ground state problem in the quantum Hamiltonian formalism, the open Wilson line operator serves as an order parameter of the $\mathbb{Z}_2[0] \times \mathbb{Z}_2[1]$ SPT phase~\cite{verresen2022Higgs}; see Ref.~\cite{2021NatPh..17..342L, kuno2023production, kuno2023bulk} for studies on MOCs using this operator and its generalization.
In this work, we use it to characterize the mixed phase rather than the `Higgs=SPT' phase in our projective model.}

When all the measurement outcomes are $+1$, then the expectation value of this operator would be a good measure of the SPT phase. 
However, the non-trivial measurement outcomes change the sign of the Wilson line operator. (Note that difference in induced states corresponding to different measurement outcomes can be accounted for by applying Pauli operators to the state with the all $+1$ outcomes.)
As such, we measure the ensemble average of the absolute value of the expectation value $\mathbb{E}|\langle W_\gamma\rangle|$.

We expect that, if we run full cycles the expectation value $\mathbb{E}|\langle W_\gamma \rangle|$ is non-zero in the regime close to the cluster state limit, $(p_J,p_K) \simeq (1,1)$. 
\rev{On the other hand,} if we stop 
\rev{after a half cycle}, 
\rev{$\mathbb{E}|\langle W_\gamma \rangle|$ is unity along the entire line $p_J = 1$ since the operator $W_e$ is always measured right before the expectation value is computed.}

\rev{\subsubsection{Comments on temporal structure and potential variants}}

\rev{
In our setup, measurement rounds alternate between the $Z$-operator round and the $X$-operator round.
Some readers may find our $ZX$-alternating temporal structure peculiar, even though such a choice is also utilized in e.g. Ref.~\cite{2020PhRvB.102i4204L,kuno2023production}.
Here, we make a few comments on this.
}

\rev{
The entanglement measures are expected to behave differently depending on whether we compute quantities after the $X$-basis round (i.e., after full cycles) or the $Z$-basis round (i.e., after an additional half cycle) as measurement itself is known to alter criticality in general (see Refs.~\cite{2023PhRvX..13b1026G,2023PhRvX..13d1042M,2023PhRvB.107x5132W,2023PhRvB.108p5120Y}).
Rather unexpectedly, for some cases (such as results in Appendix~\ref{sec:half}), we find the critical exponent extracted from TEE agrees between both scenarios.
However, we do not claim that this agreement in entanglement measures always holds and we simply keep in mind that this subtlety could exist in general.
We will see in some data below possible discrepancy in threshold probabilities and critical exponents.}

\rev{
Despite this `disadvantage,' we show that our $ZX$-alternating setup is natural from the MBQC perspective because it can be derived as a quantum circuit simulated by sequential, randomized single-qubit measurements on the Raussendorf-Bravyi-Harrington state~\cite{raussendorf2005long}, which is explained in Appendix~\ref{sec:RBH}.
From this MBQC perspective---resonating with the Bell cluster picture in Lang and B\"uchler~\cite{2020PhRvB.102i4204L}---our similar $ZX$-alternating MOC supports an underlying spacetime lattice and hence gives rise to close connection to percolation problems.
Nonetheless, one could alternatively study a setup where the choice between the $Z$- or $X$-type measurements round itself is randomized according to an additional probability (c.f. Refs.~\cite{lavasani2021topological,zhu2023structured}).
We briefly discuss such a setup in Appendix~\ref{sec:XZ-randomized}.
}

\rev{
\section{Numerical simulations}
}
\label{sec:results}
\rev{
\subsection{Simulation results}
}

Having understood the overall structure of the phase diagram, we intend to investigate the whole parameter region via numerics on the three diagnostics. 
We first describe the lattice. 
The bulk consists of vertices which are placed on integer points $(x,y) \in \{1,...,L_x\} \otimes \{1,...,L_y\} $ and edges connecting between adjacent pairs of them. 
The rough boundary consists of edges. 
In total, there are $N_q  = 3L_xL_y + L_x +L_y$ qubits. 

\subsubsection{ Topological entanglement entropy }
For the topological entanglement entropy $S_\text{topo}$, we take three bulk regions in such a way that they scale with the system size, see Fig.~\ref{fig:TEE_setup}.

\begin{figure*}
\includegraphics[width=0.8\linewidth]{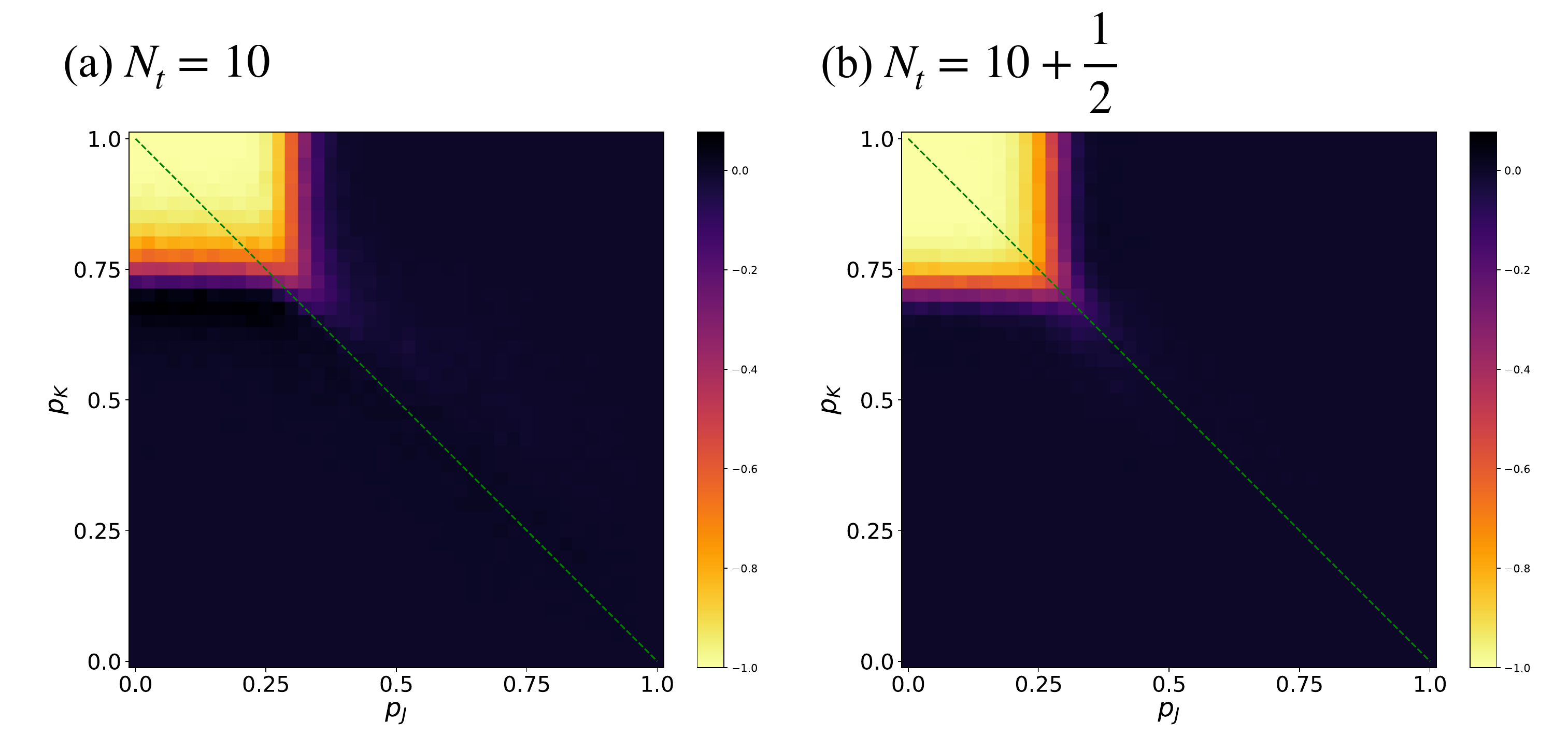}
\caption{ Topological entanglement entropy for our model with $(L_x,L_y) = (12,7)$, $N_s=1500$. (a) After 10 full cycles. (b) After 10 full cycles and a half cycle with steps (1-1) and (1-2). Both scenarios exhibit the deconfining phase at the top left corner.
\rev{We conjecture that the regions from the two different temporal slices match in the thermodynamic limit, and the bottom-right corner of the deconfining phase sits on the self-duality line $p_K=1-p_J$.}}
\label{fig:TEE_result}
\end{figure*}

We scanned the 2d parameter space with $(L_x,L_y) = (12,7)$, $N_t=10$, and $N_s=1500$ to confirm the overall picture in Fig.~\ref{fig:big-picture}.
We refer to Fig.~\ref{fig:TEE_result} for our results.
Non-zero TEE, i.e., the existence of topological order, is displayed in a finite region close to $(p_J,p_K) = (0,1)$. 
Our key observation here is that the essential structure does not differ between the $N_t=10$ full-cycle scenario versus the measuring TEE in the middle after a further half-cycle with (1-1) and (1-2). 
The TEE is a robust measure of the deconfining phase of our MOC model, in this sense.
We will see below that the precise phase boundary changes slightly as we increase the system size.

\begin{figure}
    \centering
    \includegraphics[width=0.8\linewidth]{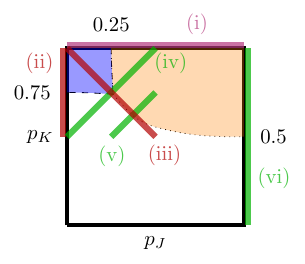}
    \caption{\rev{In  this schematic phase diagram, we show various cuts and regions in the two-dimensional parameter space where we have performed numerical simulations.}}
    \label{fig:cuts}
\end{figure}

\begin{figure*}
    \centering
    \includegraphics[width=\linewidth]{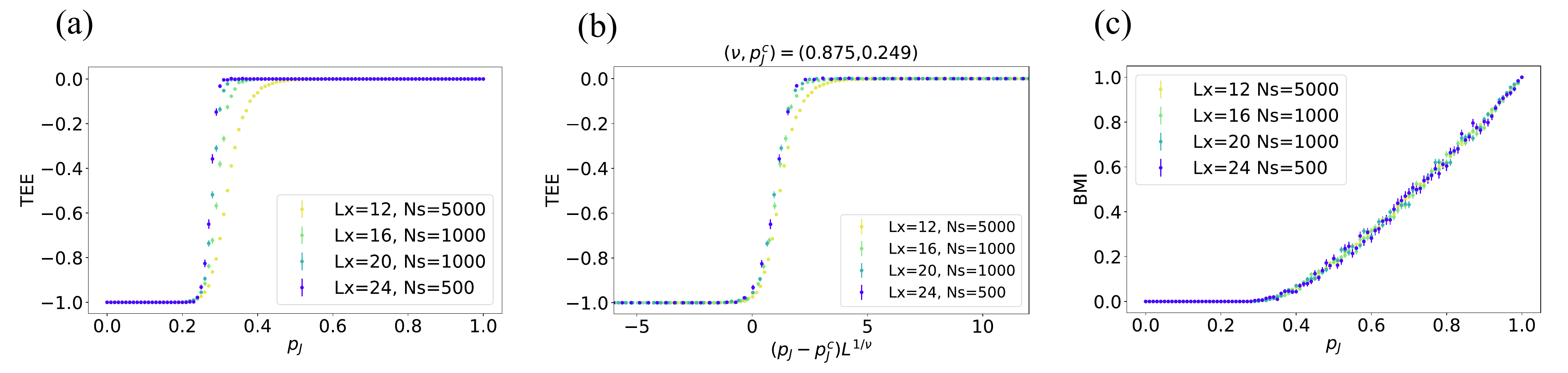}
    \caption{
    \rev{{\bf Cut (i)}: Numerical result along $p_K=1$. (a) The TEE. (b) The finite size scaling analysis for the TEE, with the best fit indicating the transition at $p^c_J = 0.25(1)$ with critical exponent $\nu = 0.87(5)$. (c) The BMI, indicating the transition at $p^c_J = 0.28(2)$. See also a companion result in Fig.~\ref{fig:cut_i_half}.}
    }
    \label{fig:cut_i}
\end{figure*}

\begin{figure*}
    \centering
    \includegraphics[width=\linewidth]{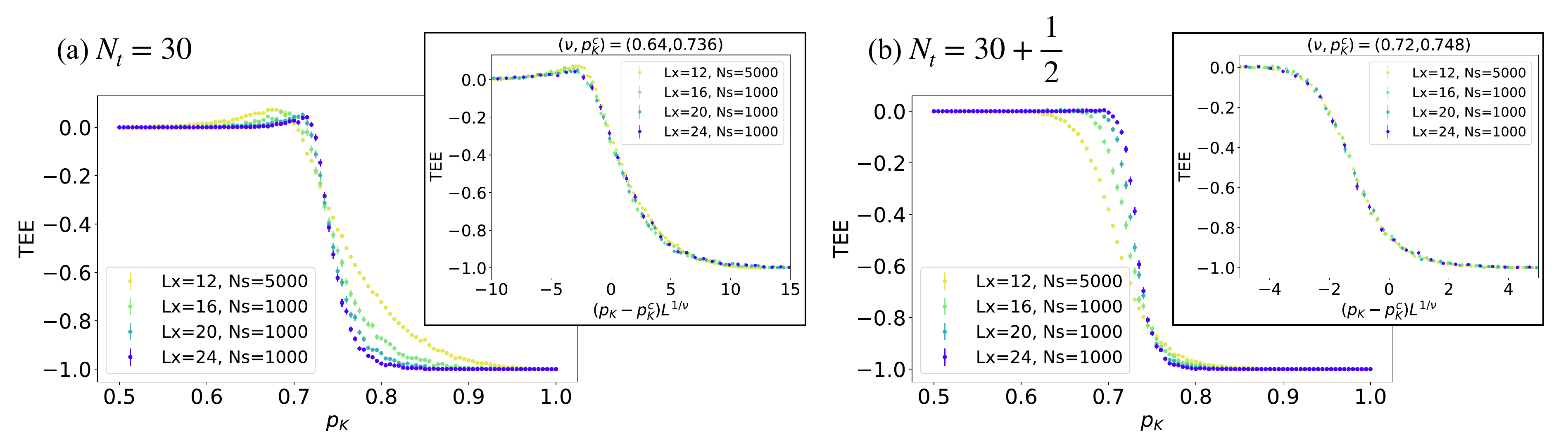}
    \caption{
    \revv{{\bf Cut (ii)}: Numerical result along $p_J=0$. (a) The TEE and its finite size scaling analysis for simulations with 30 cycles, indicating the transition at $p^c_K = 0.736(2)$ with critical exponent $\nu = 0.64(5)$. (b) The TEE and its finite size scaling analysis for simulations with 30 and a half cycles, indicating the transition at $p^c_K = 0.748(2)$ with critical exponent $\nu = 0.72(5)$.}
    }
    \label{fig:cut_ii}
\end{figure*}

\begin{figure}
    \centering
    \includegraphics[width=\linewidth]{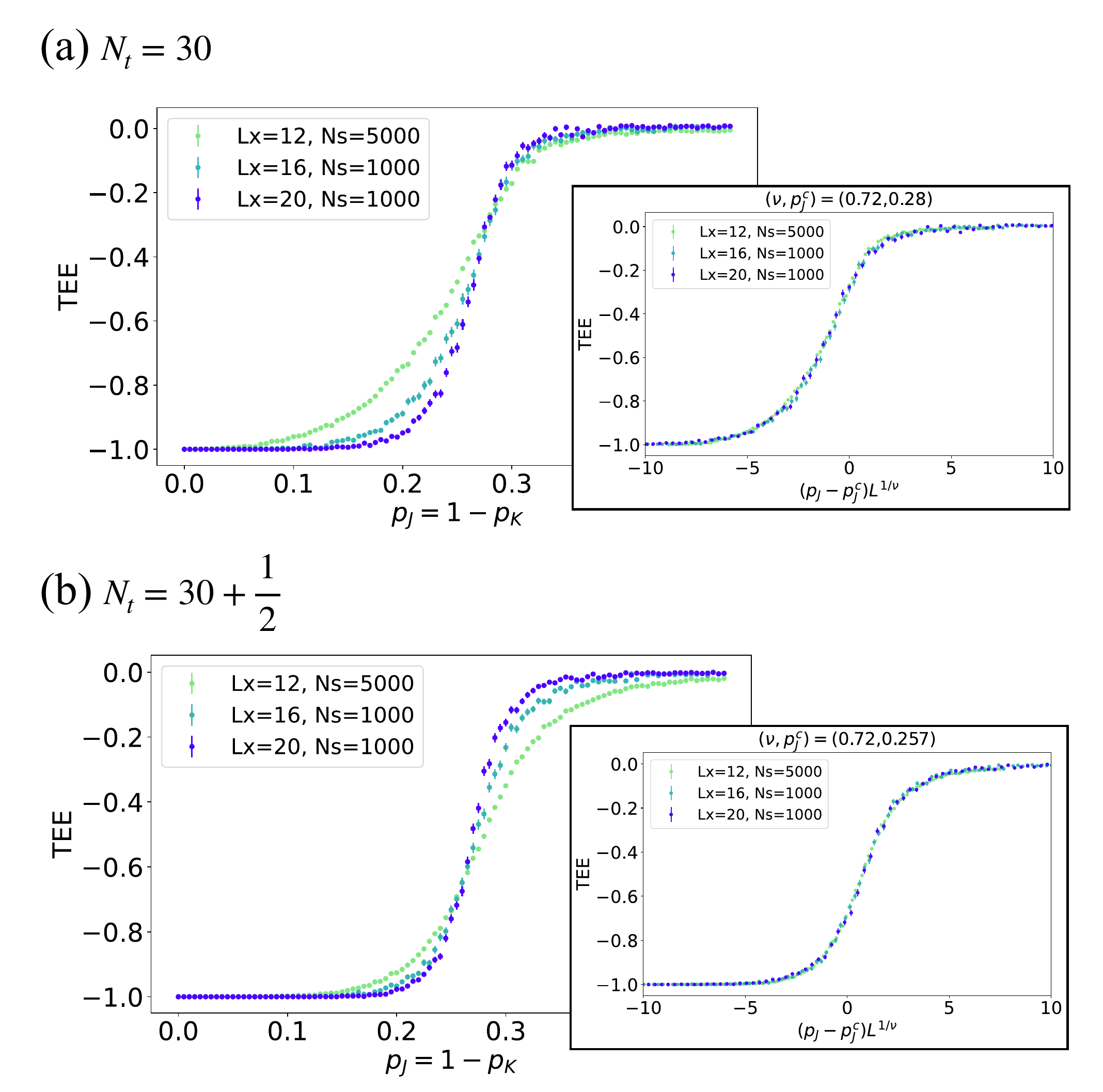}
    \caption{\rev{{\bf Cut (iii)}: an anti-diagonal scan along $p_J=1-p_K$. (a) The TEE after 30 full cycles and its finite size scaling analysis (inset). We estimate $p^c_J= 0.280(5)$ and $\nu = 0.72(5)$. (b) The TEE after 30 full cycles and a half cycleand its finite size scaling analysis (inset). We estimate $p^c_J= 0.257(3)$ and $\nu = 0.72(5)$.} }
    \label{fig:cut_iii}
\end{figure}

\begin{figure}
    \centering
    \includegraphics[width=\linewidth]{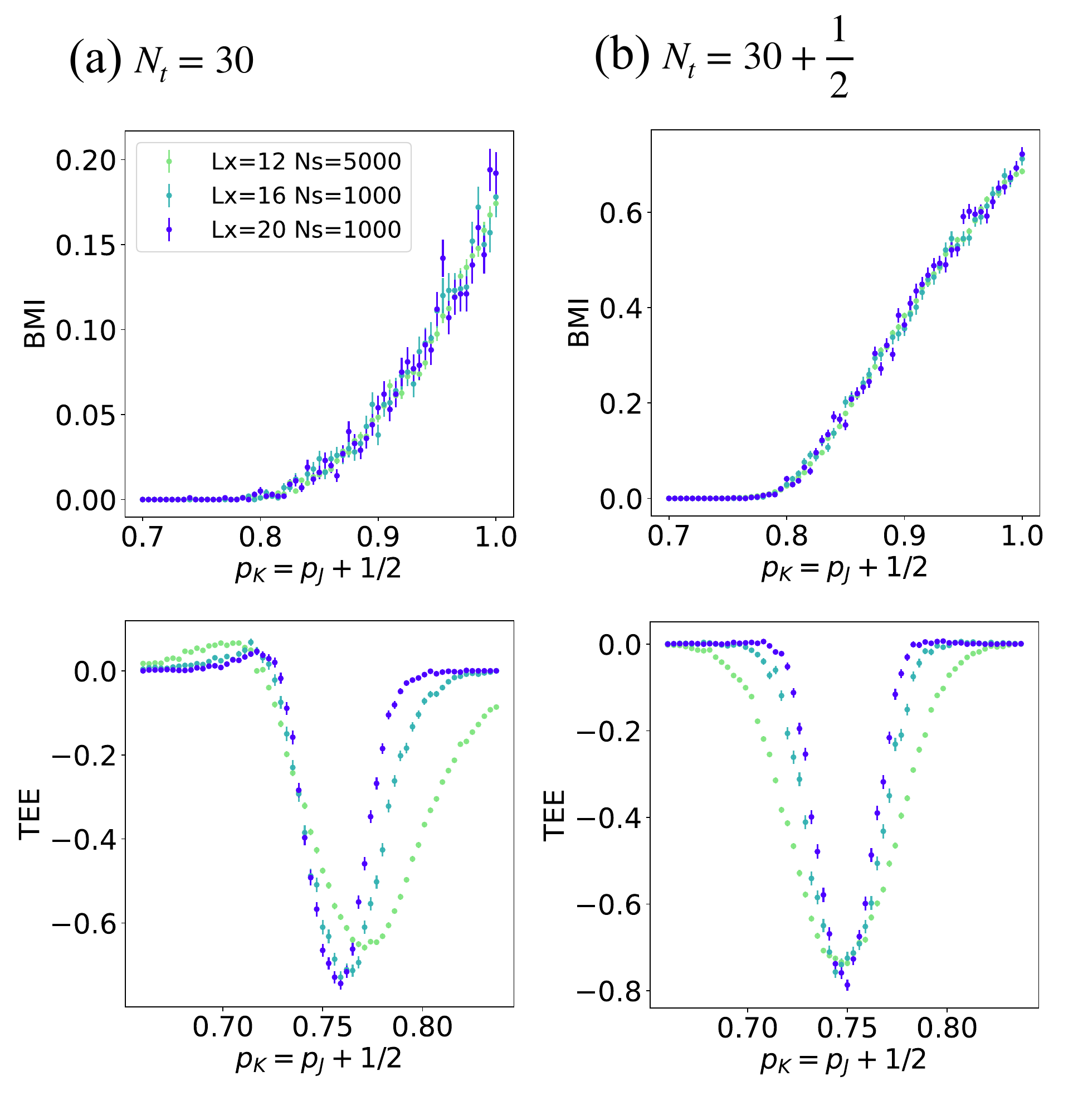}
    \caption{\rev{{\bf Cut (iv)}: a diagonal scan along $p_K= p_J + 0.5$. (a) The BMI and the TEE after 30 full cycles. (b) The BMI and the TEE after 30 full cycles and a half cycle. The BMI undergoes a transition at $p^c_K = 0.79(2)$. The TEE exhibits a dip which is a slice of the cusp in Fig.~\ref{fig:TEE_result}. The dip becomes more prominent as the system size increases, thus we expect that the cusp persists in the thermodynamic limit.}}
    \label{fig:cut_iv}
\end{figure}

We performed our numerical computation 
\rev{along various regions given in Fig.~\ref{fig:cuts} with increasing system sizes; see Figs.~\ref{fig:cut_i}, \ref{fig:cut_ii}, \ref{fig:cut_iii}, and \ref{fig:cut_iv} (see also Fig.~\ref{fig:cut_i_half}).}
We set the number of measurement rounds to be \rev{$N_t=30$}, so the spacetime volume of our system can be viewed as three dimensions.
We checked by numerics that the entanglement measures converge to their equilibrium within $N_t \sim 10$ time steps, away from transitions.

As we increase the lattice sizes, the boundary of the topological order becomes sharper.
To summarize key features, we have observed:
\begin{itemize}
\item Along (i) \rev{[Fig.~\ref{fig:cut_i} and \ref{fig:cut_i_half}]}, we see a critical point at $p^c_J\simeq 0.25$ along $p_K=1$.
\item Along (ii) \rev{[Fig.~\ref{fig:cut_ii}]}, we see a critical point at $p^c_K\simeq 0.75$ along $p_J=0$.
\item Along (iii) \rev{[Fig.~\ref{fig:cut_iii}]}, we see a transition point at $p^c_K$ which is slightly smaller than $0.75$, indicating a `cusp' of the plateau of the deconfining phase penetrating towards the center of the phase diagram.
\item Along (iv) \rev{[Fig.~\ref{fig:cut_iv}]}, we see a dip at $p_J \simeq 0.25$, consistent with the picture that a cusp exists at the right bottom corner of the deconfining phase.
\end{itemize}

\subsubsection{Boundary mutual information }

To compute the boundary mutual information $I(A:B)$, we take regions $A$ and $B$ in such a way that they contain the vertices at $(x,y)=(2,1)$ and $(x,y)=(L_x-1,1)$, respectively.
Each of the regions also contains the adjacent edge on the rough boundary.
We illustrate the setup in Fig.~\ref{fig:BMI}.
\rev{We choose $L_x\sim 2L_y$ so as to gain longer distance horizontally between two regions $A$ and $B$ in calculating the BMI $I(A,B)$.}

\begin{figure*}
\includegraphics[width=0.8\linewidth]{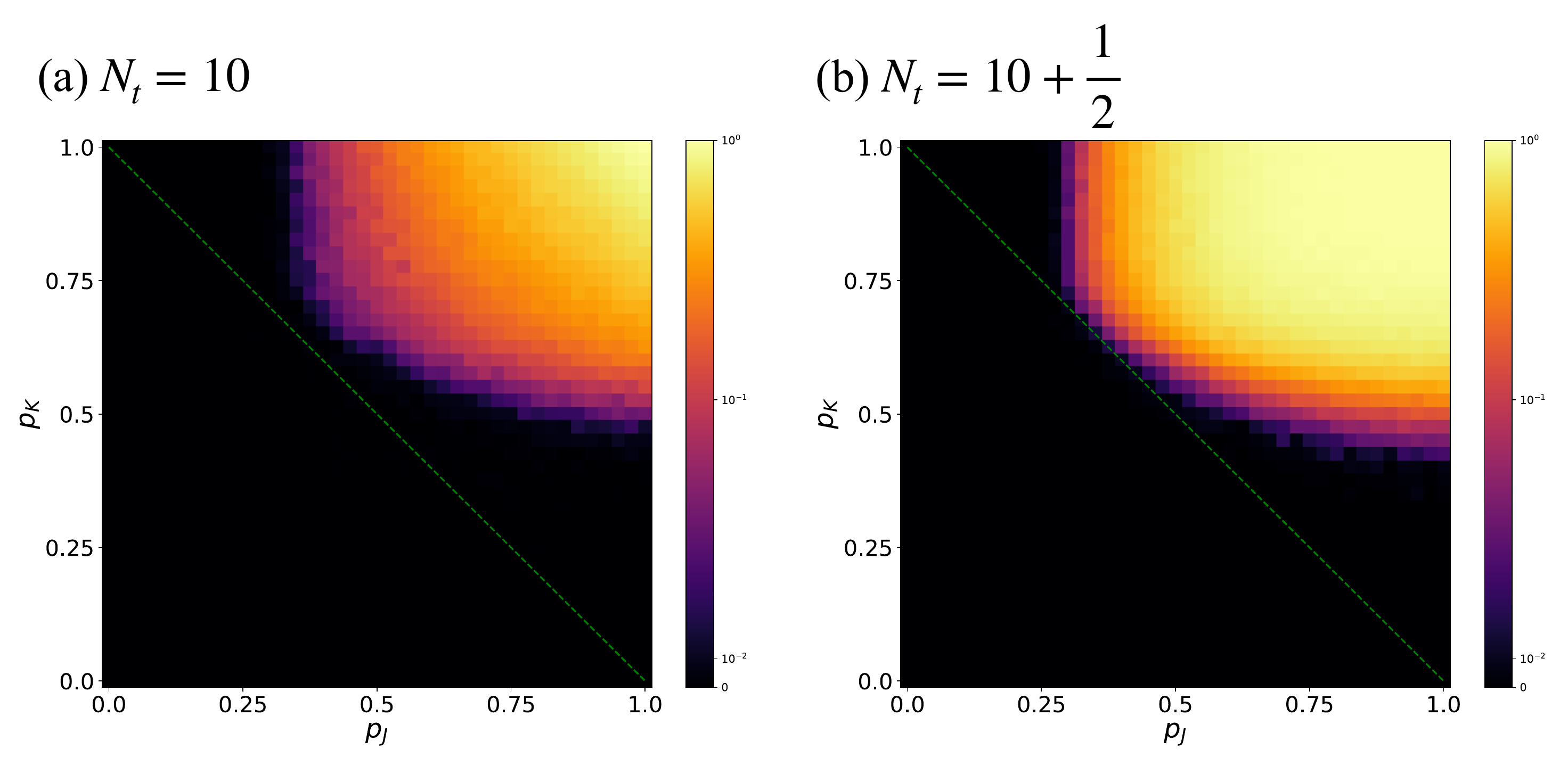}
\caption{Boundary mutual information for our model with $(L_x,L_y) = (12,7)$, $N_s=1500$. (a) After 10 full cycles. (b) After 10 full cycles and a half with steps (1-1) and (1-2). Both scenarios exhibit the Higgs=SPT phase at the finite region in the top right area.
\rev{We conjecture that the regions from the two different temporal slices match in the thermodynamic limit. We have not excluded the possibility that the Higgs=SPT phase touches the self-duality line $p_K=1-p_J$. }
}
\label{fig:BMI_result}
\end{figure*}

Similarly to the TEE, we scanned the 2d parameter space with $(L_x,L_y) = (12,7)$, $N_t=10$, and $N_s=1500$ to confirm the overall picture in Fig.~\ref{fig:big-picture}.
We refer to Fig.~\ref{fig:BMI_result} for our results.
The region with the non-zero BMI is exhibited over a finite region that contains the cluster state point $(p_J,p_K)=(1,1)$. 
We observe that the essential structure does not differ between the $N_t=10$ full-cycle scenario versus the measuring BMI in the middle after a further half-cycle with (1-1) and (1-2). Although the actual values of the BMI itself differ between the two scenarios, the onset of nonzero values agree roughly. We believe that the phase boundaries from the two scenarios match in the large system limit.
It is worth mentioning that the BMI measured at the full cycles plus a half appears sharper; see Fig.~\ref{fig:BMI_result}. 

\begin{figure*}
    \centering
    \includegraphics[width=0.7\linewidth]{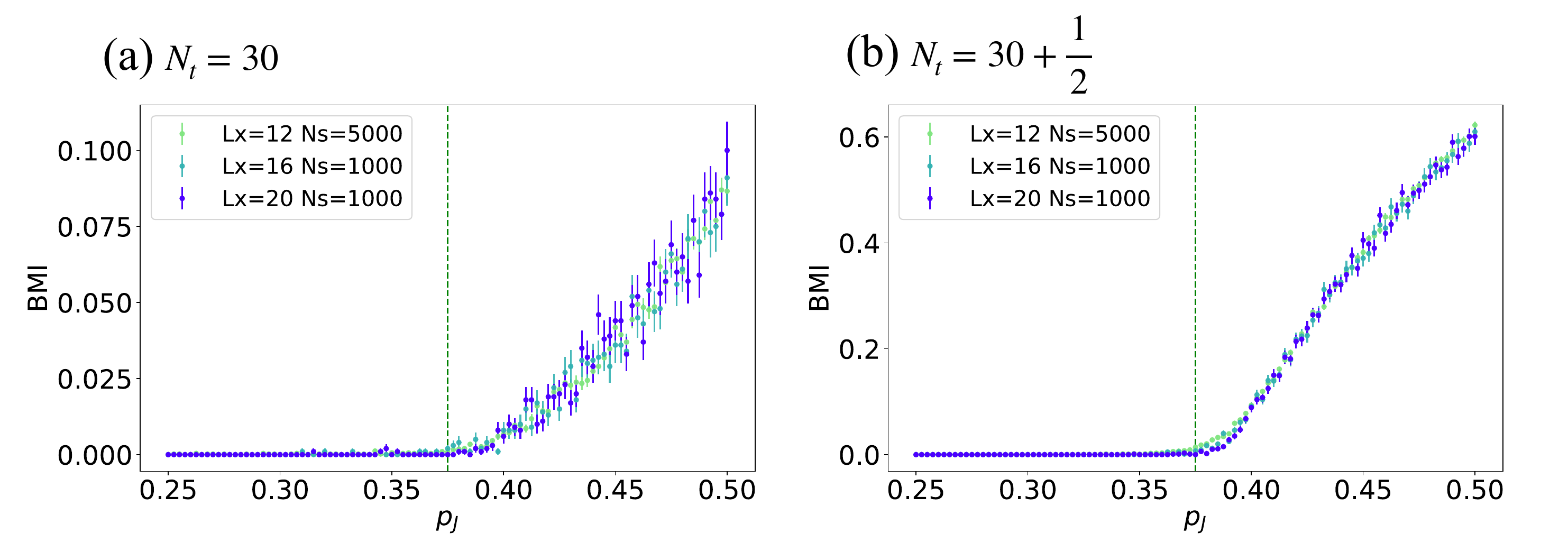}
    \caption{ \rev{{\bf Cut (v)}: the BMI along a diagonal line that crosses the self-duality line. (a) Simulation with 30 full cycles. (b) Simulation with 30 full cycles and a half cycle. Green vertical lines represent the self-duality point $p_J=0.375$. We observe transitions similar to that in Figure 3 of Ref.~\cite{2020PhRvB.102i4204L} at $p_J=0.39(1)$ which is slightly larger than the self-duality point.}}
    \label{fig:cut_v}
\end{figure*}

\begin{figure*}
    \centering
    \includegraphics[width=0.7\linewidth]{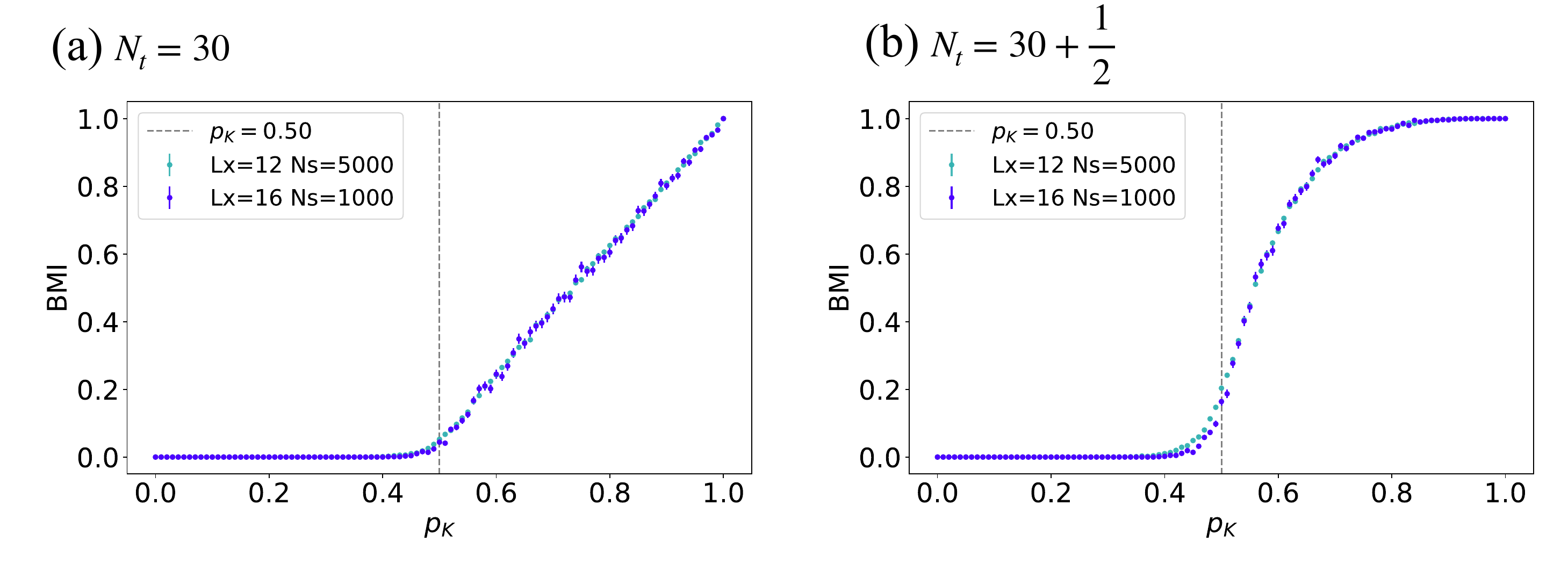}
    \caption{ \rev{{\bf Cut (vi)}: the BMI along the $p_J=1$ line. We observe the transition at $p_K=0.5$ (the dashed vertical line). (a) Simulation with 30 full cycles. (b) Simulation with 30 full cycles and a half cycle. Features of the curves, e.g. where the curvature changes, are consistent with Figure 3 of Ref.~\cite{2020PhRvB.102i4204L}, which studied the $(1+1)$d pTFI model.}}
    \label{fig:cut_vi}
\end{figure*}

We also performed our numerical computation with \rev{increased system sizes in Figs.~\ref{fig:cut_i}, \ref{fig:cut_iv}, \ref{fig:cut_v}, and \ref{fig:cut_vi}.}
From the behavior of the BMI, we have observed a second-order-like phase transitions. 
As we increase the lattice sizes, the curve appears identical for different lattice sizes to the precision of our numerics.
We have observed:
\begin{itemize}
\item Along (i) \rev{[Fig.~\ref{fig:cut_i} and \ref{fig:cut_i_half}]}, we see a critical point close to $p^c_J\simeq 0.25$ along $p_K=1$. We have not excluded a possibility that there is an intermediate phase between the deconfining phase and the Higgs=SPT phase, but we conjecture that the two boundaries match.  
\item Along (iv) \rev{[Fig.~\ref{fig:cut_iv}]}, we see a critical point at $p^c_J\simeq 0.3$. 
\rev{This can be compared with the non-zero TEE in the corresponding region.}
This is consistent with a picture that the Higgs phase and the deconfining phase are mutually exclusive.
\item  
\rev{Along (v) [Fig.~\ref{fig:cut_v}], we observe a critical point at $p^c_J=0.39(1)$ along $p_K = p_J+0.25$.}
\item Along (vi) \rev{[Fig.~\ref{fig:cut_vi}]}, we see a critical point at $p^c_K\simeq 0.5$ along $p_J=1$.
\end{itemize}

\subsubsection{A mixed phase: open Wilson line operator in the bulk}

\begin{figure}
    \includegraphics[width=1.0\linewidth]{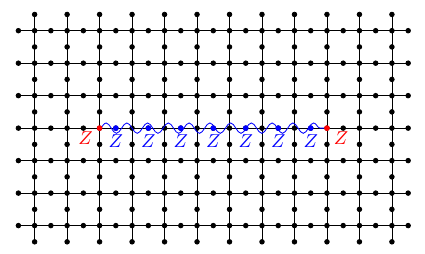}
    \caption{A setup for the open Wilson line operator $W_\gamma$.}
    \label{fig:OpenZ}
\end{figure}

Next we probe different phases appearing alternately in the protocol. We run numerical simulations for the system with $(L_x,L_y)=(12,7)$ with $N_t=10$ full cycles. We measure the open Wilson line operator inserted at the middle of the lattice extended in the $x$ direction with length $7$; see Fig.~\ref{fig:OpenZ}.

\begin{figure*}
\includegraphics[width=0.8\linewidth]{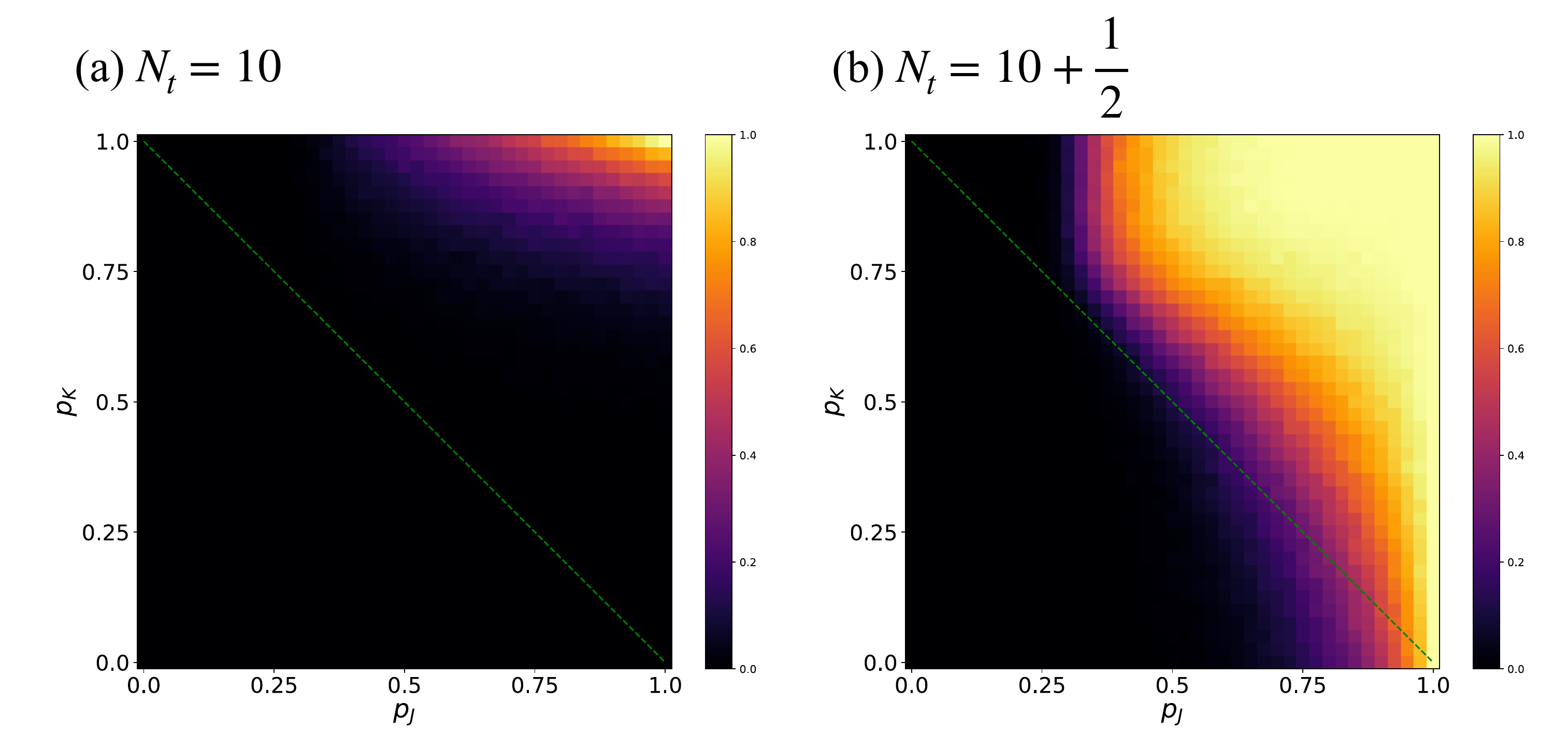}
\caption{The averaged absolute value of open Wilson line operator for our model with $(L_x,L_y) = (12,7)$, $N_s=1500$. (a) After 10 full cycles. Only concentrated region around the cluster state point shows non-zero values. (b) After 10 full cycles and a half cycle with steps (1-1) and (1-2). A large region reaching to the boundary to the deconfining phase shows non-zero expectation values. 
The two scenarios show different structures, exhibiting the mixed phase. }
\label{fig:OpenZ_result}
\end{figure*}

First, we computed the expectation value $\mathbb{E}|\langle W_\gamma \rangle|$ after the $N_t=10$ full cycles. We observe non-zero expectation values concentrated around the cluster state point $(p_J,p_K) =(1,1)$, see Fig.~\ref{fig:OpenZ_result}(a).
\rev{This is intuitive: The more away from the cluster state point, the more $X$-basis measurements occur, which remove the cluster state stabilizers $W_e$.}

\rev{
When we stop at a half cycle (after several full cycles), the measured Wilson line operator becomes surprisingly sharper and nonzero in a much wider region.
First, the region with non-zero expectation value extends all the way to $(p_J,p_K)=(1,0)$ point. This can be understood because the expectation value is unity along the line $p_J=1$ since the operator $W_e$ is always measured before the expectation value is computed. 
Second, it also reaches close to the deconfining phase, see Fig.~\ref{fig:OpenZ_result}(b).
Overall, the upper-right non-zero region in Fig.~\ref{fig:OpenZ_result}(b) matches the region with non-zero BMI in Fig.~\ref{fig:BMI_result}.
}

\begin{figure*}
    \centering
    \includegraphics[width=0.8\linewidth]{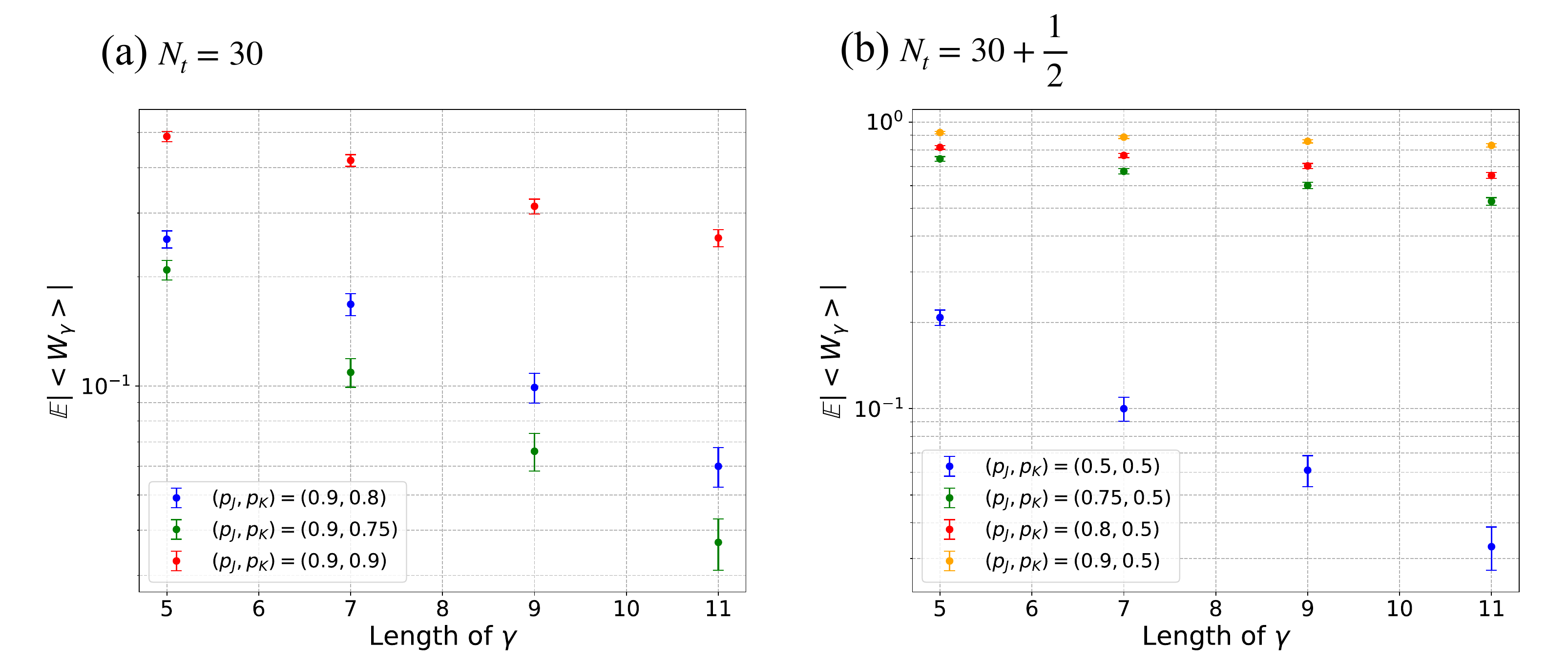}
    \caption{\rev{ Length dependence of the open Wilson line expectation value. The simulation is performed with $L_x=16$, $L_y=7$, and $N_s=1000$. (a) The expectation value measured after 30 full cycles. Away from the cluster state point $(p_J,p_K)=(1,1)$, the expectation value decays exponentially as the length of the open Wilson line increases. (b) The expectation value measured after 30 full cycles and one half cycle. Away from the cluster state line $p_J=1$, the expectation value decays exponentially as the length of the open Wilson line increases.
    Qualitative features in (a) and (b) are consistent with the corresponding diagrams in Fig.~\ref{fig:OpenZ_result}. }}
    \label{fig:length-WL}
\end{figure*}

We also computed the dependence on the length of the Wilson line operator, and found that the smaller the length is, the larger the non-zero value region becomes\rev{; see Fig.~\ref{fig:length-WL}}.

In particular, we refer to the region near $(p_J,p_K) =(1,0)$ as a mixed region, as it exhibits either more Higgs-like property or more confinement property, depending how we stop the measurement cycle. 
This is in some sense analogous to mixture of water-vapor phase in the first-order transition. 
In contrast, near the other three corners, the properties seem stable regardless of how we stop the procedure.

\subsection{Bulk phase transition, percolation, and duality}

We have obtained the transition point $p_K\simeq 0.75$ at $p_J = 0$ and $p_J\simeq 0.25$ at $p_K = 1$ from the topological entanglement entropy. 
Below, we provide some explanation of the above results from the viewpoint of percolation and duality. 

\rev{\subsubsection{Scanning (ii)}}

We first look at the transition point along the $p_J=0$ line. 
What is effectively simulated along this line is the MOC with $P^{(1)}_{B,p}$ (measuring the magnetic plaquette operator) and $P^{(3)}_{X,e}$ (measuring the electric link operator), as the vertex degrees of freedom is frozen to $X_v=1$ as a separate product state. 
This is a {\it pure-gauge} MOC model: (0) Initialize a 2d state in the product state $|+\rangle^{\otimes E}$.
(2) Associated with each plaquette, measure $B_p$ with probability $p_K$.
(3) Associated with each edge, measure $X_e$ with probability $1-p_K$, and so on.
The simulated state obeys the Gauss law constraint $A_v=1$ for all $v \in V$, where $A_v$ is the star term in the toric code,
\begin{align}
A_v 
&= \prod_{e \supset v} X_e 
=
\raisebox{-20pt}{\begin{tikzpicture}
\draw[-,black!30,line width=1.0] (-0.5,-0.0) -- (0.5,-0.0);
\draw[-,black!30,line width=1.0] (0.0,-0.5) -- (0.0,0.5);
\node at (-0.5,0.0) {$X$};
\node at (0.5,0.0) {$X$};
\node at (0.0,0.5) {$X$};
\node at (0.0,-0.5) {$X$};
\end{tikzpicture} }
\quad \forall v \in V . 
\end{align}
In Appendix~\ref{sec:RBH}, generalizing the relation between the bond percolation and pTFI, we argue that a phase transition in the pure gauge MOC can be described by a homological percolation~\cite{aizenman1983sharp, 2020arXiv201111903D} of surfaces in the 3d square lattice.
Indeed, the surface percolation in 3d lattice is dual to the bond percolation with a relation in thresholds $\tilde{p}^\text{bond}_c + \tilde{p}^\text{surface}_c=1$ (see~\cite{aizenman1983sharp, 2020arXiv201111903D}).
According to Refs.~\cite{van1997percolation, lorenz1998precise}, $\tilde{p}^{\rm bond}_c = 0.2488126(5)$, so 
\rev{a}
phase transition is 
\rev{expected at $p_K\simeq 0.75$.}

\rev{We show our numerical results for cut (ii) in Fig.~\ref{fig:cut_ii}.
We observe the transition at $p_K\simeq 0.74(1)$. 
On contrary, depending on whether we compute the topological entanglement entropy after full cycles or after an additional half cycle, the critical exponents seem to differ somewhat slightly.
Given our argument above, the criticality would be described by some three-dimensional conformal field theory, but we are not aware of existing values for critical exponents for the homological percolation problem in the literature which we can compare with.}

\rev{\subsubsection{Scanning (i)}}

For the transition along the $p_K=1$ line, we use a duality to consider an equivalent model. 
Consider a map 
\begin{align}
\mathbf{\Sigma} = \langle+|^{\otimes V} \prod_{\langle e,v\rangle} CX_{v,e} ,
\end{align}
where $\prod_{\langle e,v \rangle}$ is the product over adjacent pairs of vertices and edges and $CX_{a,b}$ is the controlled-$X$ operator, $CX_{a,b} = |0\rangle\langle0|_a \otimes I_b + |1\rangle\langle1|_a \otimes X_b $.
Then, the measurement bases on the $p_K=1$ line are mapped as follows:
\begin{align}
\mathbf{\Sigma} \cdot 
\begin{pmatrix}
X_v \\
W_e \\
B_p
\end{pmatrix}
= 
\begin{pmatrix}
A_v \\
Z_e \\
B_p
\end{pmatrix}
\cdot \mathbf{\Sigma} .
\end{align}
Thus the projective gauge-Higgs MOC model with bases $(X_v,W_e,B_p)$ and probability assignment $(1-p_J,p_J,1)$ is equivalent to that with $(A_v,Z_e,B_p)$ and probability assignment $(1-p_J,p_J,1)$.
By further transforming them using the Hadamard transform and viewing the terms as those in the dual lattice, we have an MOC model with $(B_{p^*},X_{e^*},A_{v^*})$ and probability assignment $(1-p_J,p_J,1)$.  
It is the pure-gauge MOC model that we described above (with a different initial state), and thus the transition point is expected at $1-p_J=\tilde{p}^\text{surface}_c$, so $p_J=\tilde{p}^\text{bond}_c$.
This is consistent with our critical point, $p_J \simeq 0.25$.

We note that there is another map that we can make use of.
The map is simply a projector 
\begin{align}
\mathbf{\Delta} = \langle 0|^{\otimes E} .
\end{align}
Essentially, the operator $\mathbf{\Delta} $ forces to solve for the Gauss law constraints in the gauge theory and leaves us a model without gauge redundancy but with the $\mathbb{Z}_2$ global symmetry.
We get 
\begin{align}
\mathbf{\Delta} \cdot 
\begin{pmatrix}
X_v \\
W_e \\
B_p
\end{pmatrix}
= 
\begin{pmatrix}
X_v \\
L_e \\
I
\end{pmatrix}
\cdot \mathbf{\Delta} 
\end{align}
with $L_e = \prod_{v \subset e} Z_v$ being the Ising interaction,
\begin{align}
L_e &= \prod_{v \subset e} Z_v
=\raisebox{-10pt}{\begin{tikzpicture}
\draw[-,black!30,line width=1.0] (-0.4,0.0) -- (1.4,0.0);
\draw[-,black!30,line width=1.0] (0.0,-0.4) -- (0.0,0.4);
\draw[-,black!30,line width=1.0] (1.0,-0.4) -- (1.0,0.4);
\node at (1.0,0.0) {$Z$};
\node at (0.0,0.0) {$Z$};
\end{tikzpicture} } .
\end{align}
Therefore, the projective gauge-Higgs MOC model with measurement bases 
$(W_e, X_v, B_p)$ and probability assignment $(p_J, 1-p_J, 1)$ is equivalent to that with $(X_v, L_e)$ and probability assignment $(p_J, 1-p_J)$. The latter model is the $(2+1)$d pTFI model, which was also studied by Lang and B\"{u}chler.
Their numerical result and the percolation picture indicated that the transition point is at the 3d bond percolation threshold $\tilde{p}^{\rm 3d}_c = 0.2488126(5)$ for the square lattice.
\rev{Our finite size scaling analysis in Fig.~\ref{fig:cut_i}(b) indicates $p^c_J=0.25(1)$, and further we estimate the critical exponent as $\nu=0.87(5)$, which is consistent with $\nu = 7/8$~(see e.g. \cite{lorenz1998precise}). }

\rev{We note that a map to a bond percolation model was constructed also by Lavasani et al. in Ref.~\cite{lavasani2021topological}, where a `pure gauge MOC' was studied extensively. 
We stress that our pure gauge MOC model is different from theirs in detail; their circuit does not have the $ZX$-alternating temporal structure as in ours. As such, we obtain a different threshold probability.}

\subsubsection{ \revv{Cut (i) vs Cut (ii)} }

\revv{We point out an open question regarding our numerical result. 
We notice that we obtained different critical exponents in Figs.~\ref{fig:cut_i} and \ref{fig:cut_ii}.
On contrary,  Cut (i) and (ii) seem to be directly related to each other. 
First, we showed above that the projective measurement pattern along Cut (i) is mapped via $\mathbf{\Sigma}$ to the $(2+1)$d pure gauge MOC model. 
Second, the model along Cut (ii) is nothing but the $(2+1)$d pure gauge MOC model.
The tension between the analytical understanding versus the numerical result may be attributed to either numerical insufficiency (although the data collapses are seemingly perfect) or something more subtle. 
Regarding the latter scenario, we notice that the map $\mathbf{\Sigma}$ was recently used in Ref.~\cite{2023arXiv231116235X} to study entanglement spectrum in the gauge-Higgs model and its dual, and they show that there is indeed difference in entanglement spectrum. 
Although we don't have a concrete argument, we leave that the duality map may cause subtle effects in entanglement criticality as a possibility of explaining the discrepancy.
}

\subsection{Boundary phase transition and percolation via screening}

Next, we unmask the physics behind the boundary phase transitions. 
Along the $p_K=1$ line, we had a transition at 
\rev{$p_J = 0.28(2)$ as indicated in Fig.~\ref{fig:cut_i}(c)},
consistent with the fact that an SPT phase is short-range entangled, so it cannot coexist with the deconfining phase. 

How about the $p_J=1$ line, where we observed a transition at $p_K\simeq 0.5$?
Below, we claim that the effective theory is a boundary $(1+1)$d pTFI model and the bulk is effectively decoupled, so the critical point is that of the $(1+1)$d pTFI in disguise.

To derive the effective theory, we use the cluster state entangler 
\begin{align}
\mathbf{ \mathcal{C} }
= \prod_{\langle e,v \rangle } CX_{v,e} . 
\end{align}
Then the measurement bases relevant in the $p_J=1$ line are transformed as
\begin{align}
\raisebox{-10pt}{\begin{tikzpicture}
\draw[-,black!30,line width=1.0] (-0.4,0.0) -- (1.4,0.0);
\draw[-,black!30,line width=1.0] (0.0,-0.4) -- (0.0,0.4);
\draw[-,black!30,line width=1.0] (1.0,-0.4) -- (1.0,0.4);
\node at (1.0,0.0) {$Z$};
\node at (0.5,0.0) {$Z$};
\node at (0.0,0.0) {$Z$};
\end{tikzpicture} }
&\overset{
\mathbf{ \mathcal{C} } }{\longleftrightarrow}
\raisebox{-10pt}{\begin{tikzpicture}
\draw[-,black!30,line width=1.0] (-0.4,0.0) -- (1.4,0.0);
\draw[-,black!30,line width=1.0] (0.0,-0.4) -- (0.0,0.4);
\draw[-,black!30,line width=1.0] (1.0,-0.4) -- (1.0,0.4);
\node at (0.5,0.0) {$Z$};
\end{tikzpicture} } \quad e \in E_{B} ,  \label{eq:Z-measure-dual}\\ 
\raisebox{-20pt}{\begin{tikzpicture}
\draw[-,black!30,line width=1.0] (-0.5,-0.5) -- (0.5,-0.5) -- (0.5,0.5) -- (-0.5,0.5) -- (-0.5,-0.5);
\node at (-0.5,0.05) {$Z$};
\node at (0.5,0.05) {$Z$};
\node at (0.0,0.53) {$Z$};
\node at (0.0,-0.45) {$Z$};
\end{tikzpicture} }
&\overset{
\mathbf{ \mathcal{C} } }{\longleftrightarrow}
\raisebox{-20pt}{\begin{tikzpicture}
\draw[-,black!30,line width=1.0] (-0.5,-0.5) -- (0.5,-0.5) -- (0.5,0.5) -- (-0.5,0.5) -- (-0.5,-0.5);
\node at (-0.5,0.05) {$Z$};
\node at (0.5,0.05) {$Z$};
\node at (0.0,0.53) {$Z$};
\node at (0.0,-0.45) {$Z$};
\end{tikzpicture} } \quad p \in P_{B}  ,\\
\raisebox{-10pt}{\begin{tikzpicture}
\draw[-,black!30,line width=1.0] (-0.4,0.0) -- (1.4,0.0);
\draw[-,black!30,line width=1.0] (0.0,-0.4) -- (0.0,0.4);
\draw[-,black!30,line width=1.0] (1.0,-0.4) -- (1.0,0.4);
\node at (0.5,0.0) {$X$};
\end{tikzpicture} }
&\overset{
\mathbf{ \mathcal{C} } }{\longleftrightarrow}
\raisebox{-10pt}{\begin{tikzpicture}
\draw[-,black!30,line width=1.0] (-0.4,0.0) -- (1.4,0.0);
\draw[-,black!30,line width=1.0] (0.0,-0.4) -- (0.0,0.4);
\draw[-,black!30,line width=1.0] (1.0,-0.4) -- (1.0,0.4);
\node at (0.5,0.0) {$X$};
\end{tikzpicture} } \quad e \in E_{B\partial} .
\end{align}
The measurement with the basis \eqref{eq:Z-measure-dual} occurs with the unit probability in the bulk (due to $p_J=1$ in the original theory), so the $Z$ measurement round is trivial in the sense that the plaquette operator $B_p$ is locked to $B_p=1$. 
Therefore, the bulk system in this basis experiences no entangling dynamics; it is simply a cycle of (1) reset to $Z_e=1$ (2) measure $X_e$ with probability $1-p_K$.

The plaquette opeator on the boundary, on the other hand, is non-trivial. We get 
\begin{align}
\raisebox{-20pt}{\begin{tikzpicture}
\draw[-,black!30,line width=1.0] 
(-0.5,0.5)--(-0.5,-0.5)--(0.5,-0.5)--(0.5,0.5);
\draw[-,dotted,black!80] (-1.0,0.0) -- (1.0,0.0);
\node at (-0.5,0.05) {$Z$};
\node at (0.5,0.05) {$Z$};
\node at (0.0,-0.45) {$Z$};
\end{tikzpicture} }
&\overset{
\mathbf{ \mathcal{C} } }{\longleftrightarrow}
\raisebox{-20pt}{\begin{tikzpicture}
\draw[-,black!30,line width=1.0](-0.5,0.5)--(-0.5,-0.5)--(0.5,-0.5)--(0.5,0.5);
\draw[-,dotted,black!80] (-1.0,0.0) -- (1.0,0.0);
\node at (-0.5,0.05) {$Z$};
\node at (0.5,0.05) {$Z$};
\node at (0.0,-0.45) {$Z$};
\end{tikzpicture} } 
=
\raisebox{-20pt}{\begin{tikzpicture}
\draw[-,black!30,line width=1.0](-0.5,0.5)--(-0.5,-0.5)--(0.5,-0.5)--(0.5,0.5);
\draw[-,dotted,black!80] (-1.0,0.0) -- (1.0,0.0);
\node at (-0.5,0.05) {$Z$};
\node at (0.5,0.05) {$Z$};
\node at (0.0,-0.45) {$+1$};
\end{tikzpicture} }  \label{eq:ZZ-dual}
\end{align}
for $p \in P_{\partial}$. In the equality, we used the condition $Z_e=1$ ($e \in E_B$) from the bulk measurement with unit probability.
The measurement with the basis~\eqref{eq:ZZ-dual} occurs with the probability $p_K$, and that with the basis $X_e$ $(e \in E_\partial)$ takes place with the probability $1-p_K$.
Hence, we conclude that the effective theory related to the projective gauge-Higgs MOC model at $p_J=1$ via $\mathbf{\mathcal{C}}$ is:
\begin{itemize}
\item the decoupled bulk system with a cycle of single qubit measurement of $X_e$ and resetting to $Z_e=1$, 
\item the boundary $(1+1)$d MOC circuit with the Ising basis~\eqref{eq:ZZ-dual} with probability $p_K$ and the single-qubit basis $X_e$ with probability $1-p_K$, i.e., the 1d pTFI model.
\end{itemize}

\begin{figure*}
    \centering
    \includegraphics[width=0.7\linewidth]{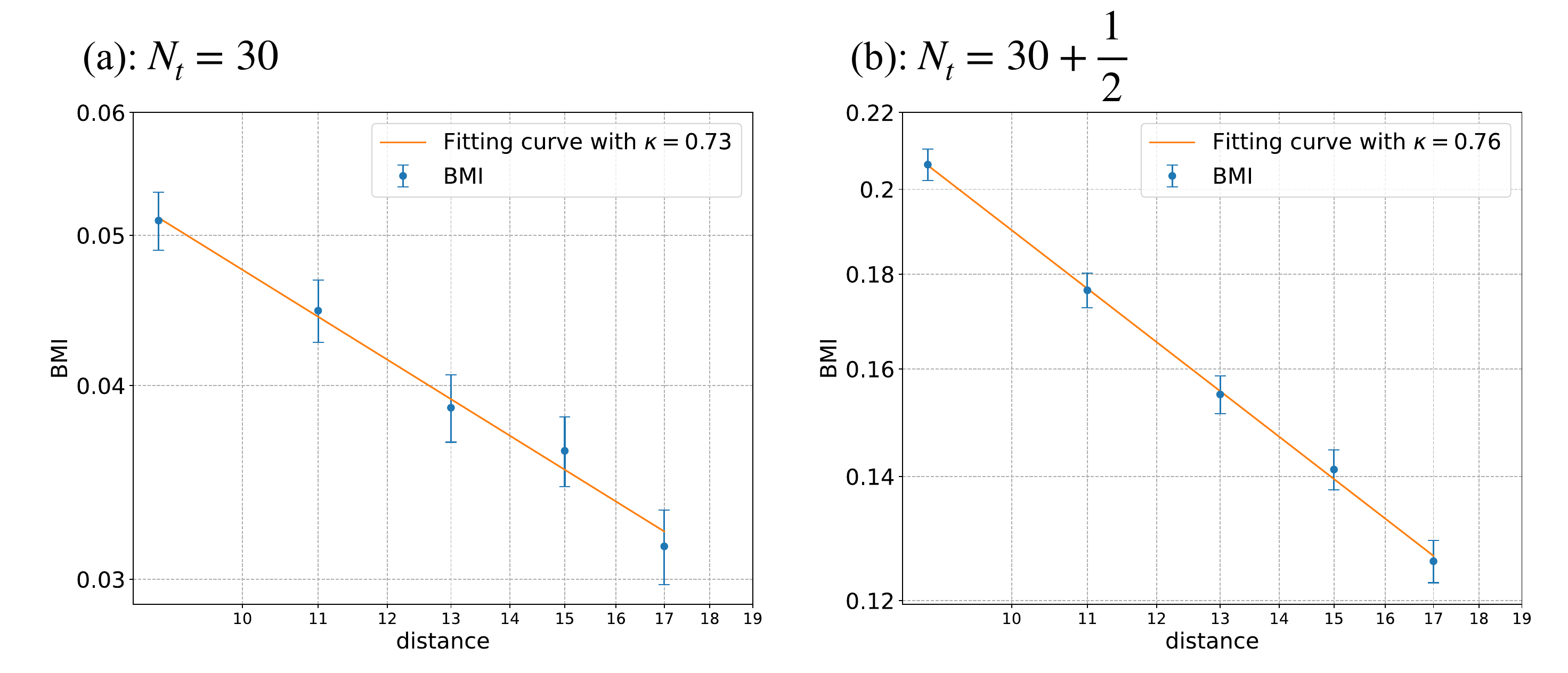}
    \caption{ \revv{The boundary mutual information at $(p_J, p_K) = (1, 0.5)$ as a function of the distance between subsystems $A$, $B$.  The curve shows an algebraic decay $I(A,B) \simeq \alpha/d(A,B)^\kappa$. The data are shown for simulations with $L_x =20$, $L_y=10$,  and $N_s = 10000$. (a) After 30 cycles. The fitting shows $\kappa = 0.73(4)$.  (b) After 30 and a half cycles. The fitting shows $\kappa = 0.76(1)$. }}
    \label{fig:pTFI_kappa}
\end{figure*}

\rev{
We support the above map to the $(1+1)$d pTFI model by the following two numerical results.
First, the critical measurement rate of the $(1+1)$d pTFI is $0.5$ (the percolation threshold on the 2d square lattice), and our result of critical $p_K \simeq 0.5$ along $p_J=1$ is consistent.
Second, the work~\cite{2020PhRvB.102i4204L} reveals the criticality of the $(1+1)$d pTFI model to be a conformal field theory (CFT) whose vertex operator has a scaling dimension such that one can determine $I(i,j) \simeq \alpha/|i-j|^\kappa$ for two points on the 1d periodic chain with $\alpha$ a non-universal coefficient and $\kappa=2/3$ the critical exponent.
In Fig.~\ref{fig:pTFI_kappa}, we show a plot of the boundary mutual information over varied distances between subsystems, denoted as $d(A,B)$.
Via fitting to data from simulating the  $(L_x,L_y)=(20,10)$ system,  whose 1d boundary system is still small, we obtain the exponent $\kappa = 0.73(4)$ and it does not contradict with the CFT prediction $\kappa=2/3$.
Larger sizes would likely improve the agreement. 
}

\begin{figure*}
    \centering
    \includegraphics[width=0.7\linewidth]{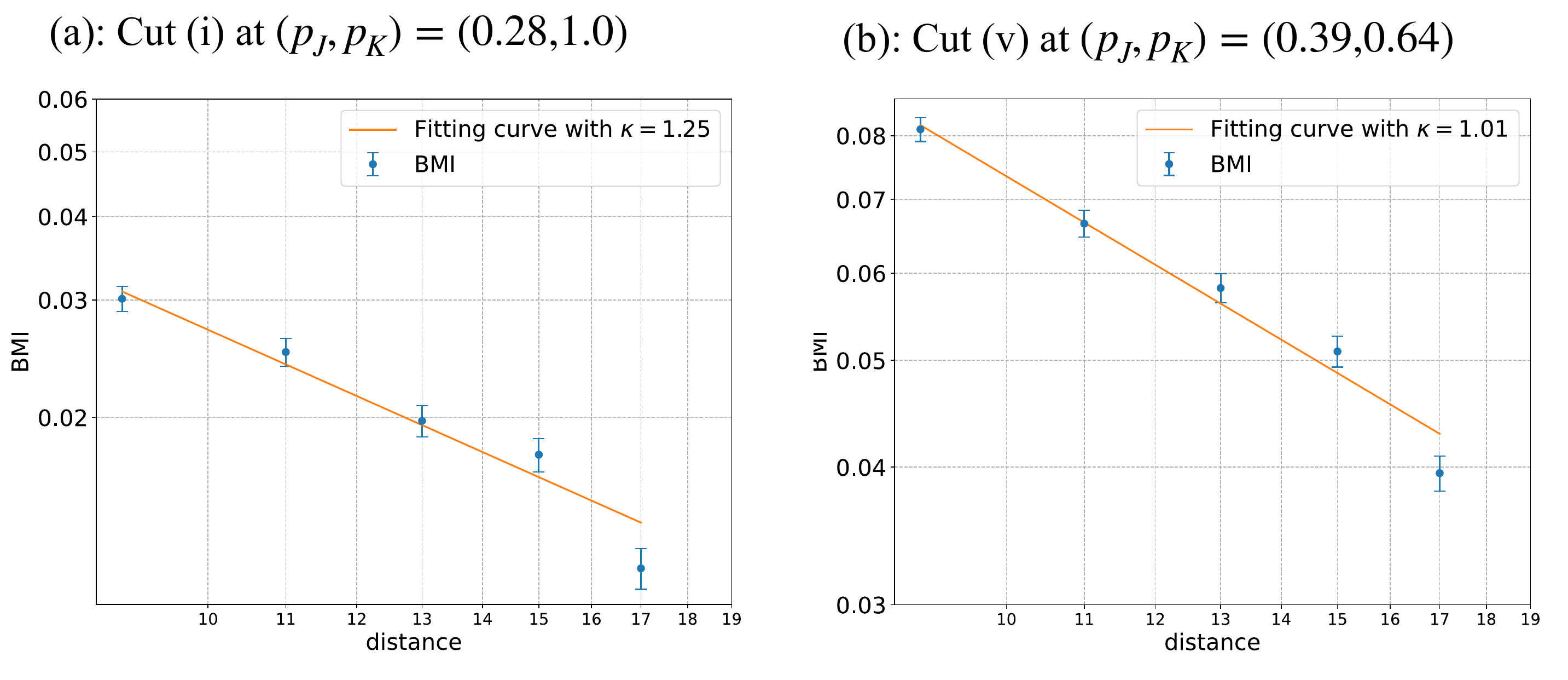}
    \caption{ \revv{ (a) The boundary mutual information as a function of distance shows an algebraic decay. (a) At $(p_J, p_K) = (0.28, 1.0)$, the curve is fitted with $\kappa = 1.25(15)$. The data are shown for simulations with $L_x =20$, $L_y=10$, $N_t=30+1/2$, and $N_s = 17000$. (b) At $(p_J, p_K) = (0.39, 0.64)$ the curve decay with $\kappa = 1.01(9)$. The data are shown for simulations with $L_x =20$, $L_y=10$, $N_t=30+1/2$, and $N_s = 18000$.  }}
    \label{fig:pTFI_kappa_i_and_v}
\end{figure*}

\revv{To demonstrate the boundary criticality away from the above $(1+1)$d pTFI limit, we initially estimate critical points as probability coordinates where the BMI curve start to rise along cuts. Despite that BMI has not been useful to perform finite size scaling analysis, just as above, we instead show criticalities by demonstrating algebraic decay of BMI as a function of distance, see Fig.~\ref{fig:pTFI_kappa_i_and_v}. 
The fitting indicates the following. 
First, it is likely that the critical theory shifts as we move along the boundary entanglement transition from the $(1+1)$d pTFI limit at $(p_J,p_K)=(1.0,0.5)$ to the $(2+1)$d pTFI limit at $(p_J,p_K)=(0.28,1.0)$. We indeed see the critical exponent $\kappa$ shifts along the transition line. Second, there remains the possibility that, along the cut (i), there is a finite region $0.25 < p_J < 0.28$ which does not belong either to the SPT phase or to the deconfining phase. Further numerical analysis with larger system sizes is again desired to confirm this point.}

\rev{\subsection{Self-duality line}}

\rev{Finally, we give some remarks on the fascinating feature of the gauge-Higgs model; namely, the self-duality.
We considered measurements of
$(B_p, W_e , X_e , X_v)$ with the probability profile $(p_K, p_J, 1-p_K, 1-p_J)$.
The duality map (see e.g.~\cite{Okuda-Parayil-Mana-Sukeno} for an operator that realizes this map) transforms this model into another model of measurements $(X_v, X_e , W_e , B_p)$ with the probability profile $(p_K, p_J, 1-p_K, 1-p_J)$ on the dual 2d square lattice.
On a line $p_K = 1- p_J$, the original model and the dual model become identical up to differences in boundaries: Roughly, the FS-MOC model is self dual on this line.
}

\rev{In Figures, we draw the self-duality line in our FS-MOC model.
From Fig.~\ref{fig:TEE_result}, we conjecture that the vertical and horizontal phase boundaries of the deconfining phase merge into the self-duality line.
In Fig.~\ref{fig:BMI_result}, it is elusive whether or not the Higgs=SPT phase meet the self-duality line. 
A scan along cut (v) in Fig.~\ref{fig:cut_v} shows that the Higgs=SPT phase does not meet the self-duality line along this cut; another scan along cut (iv) in Fig.~\ref{fig:cut_iv} also shows the same. 
However, we have not excluded the possibility that the Higgs=SPT phase touches the self-duality line in between cuts (iv) and (v), i.e., $0.25<p_J<0.375$.
We leave in-depth studies of this interesting region, where the three topological phases and the self-duality line may merge, as a future exploration.}

\revv{We also performed numerical simulation with Cut (iii) along the self-duality line as shown in Fig.~\ref{fig:cut_iii}. There, finite size scaling with TEE shows a critical exponent $\nu = 0.72(5)$. In the literature, on the other hand, the standard statistical mechanical gauge-Higgs model has been studied intensively around the multi-critical point. We notice that work~\cite{PhysRevX.11.041008} obtained $\nu = 0.669(4)$, which does not contradict with our estimation. Intriguingly, their numerical method is based on mapping of the statistical mechanical partition function to a percolation problem of anyon worldlines. It would be interesting to relate their mapping to our homological percolation picture in the MBQC method partly explained in Appendix~\ref{sec:RBH}.}

\section{Conclusions and discussion} \label{sec:conc-diss}

In this work, we considered an MOC model inspired by the Fradkin-Shenker model.
We computed the topological entanglement entropy to diagnose the topological order in the simulated state, and we found a clear phase transition across varied measurement rates.
Further, motivated by the recent finding of the boundary symmetry-breaking order, we diagnosed such a boundary long-range order by mutual information. 
The structure of our phase diagram is consistent with the one obtained from the corresponding models via the numerical DMRG method~\cite{verresen2022Higgs}, for example.

The above two topological phases, either intrinsic or symmetry-protected, are robust entanglement orders of our MO Fradkin-Shenker model. 
We also showed that our MO circuit exhibits a mixed phase, which corresponds to the parameter regime with frustrated terms in the quantum Hamiltonian picture. 
The bulk open Wilson line operator detects states alternating over measurement rounds, wherein we saw a continuity between the Higgs limit and the confining limit as well as a clear separation from the deconfining phase.

Our results suggest that it will be possible to realize the Higgs=SPT phenomenon on quantum devices in the near future.
Recent progress in quantum technology has allowed us to program mid-circuit measurements in quantum devices, such as superconducting or trapped-ion processors. 
Some of other devices that allow mid-circuit measurements, such as reconfigurable atom arrays, may not support the universal quantum computation, but they would still be useful for realizing MOC models.
Although quantum operations are still noisy as of present, it is worth noting that one can translate an MOC model into an MBQC model, as discussed in Appendix~\ref{sec:RBH}.
There, the resource state (in one dimension higher) can be prepared with a constant-depth unitary circuit, and
an MOC circuit can be simulated by Pauli-basis measurements in the bulk, with the output state induced at the boundary of the resource state. 
Crucially, one can perform the bulk measurements {\it simultaneously} as we do not need to implement $SU(2)$ gates, which will reduce the circuit complexity to {\it constant}.
Hence, it is speculated that studies of MOC models on experimental platforms are not so far-fetched within the near-future technology.
\rev{A related idea was explored by Liu et al.~\cite{liu2022measurement}.
In Ref.~\cite{Okuda-Parayil-Mana-Sukeno}, one of us constructed a resource state for the Fradkin-Shenker model in the context of Hamiltonian quantum simulation. 
One can simulate our FS-MOC model by  appropriately choosing the bases used to measure the resource state, generalizing Appendix~\ref{sec:RBH}, where we explain how to obtain the pure-gauge MOC from measurements on the Raussendorf-Bravyi-Harrington state.}

One intriguing open question is whether there is a multi-critical point in the MOC phase diagram as found in the 3d Euclidean lattice field theory of the Fradkin-Shenker model~\cite{tupitsyn2010topological}. 
Answering this question, of course, requires much more numerical effort. 
It would also be useful to understand the relation between the ordinary 3d Euclidean path integral and the replica manifold for entanglement entropy of MOCs, as well as the role of the duality in entanglement spectrum~\cite{2023arXiv231116235X}.
We hope to advance our study in this direction in the future.

\smallskip
\noindent
{\it Note added:} Recently, a paper appeared on arXiv~\cite{linsel2024percolation}, where authors made the connection between the gauge-Higgs model and percolation. We deduced a connection to percolation based on an MBQC picture.
\rev{After the submission of this work (arXiv: 2402.11738) to arXiv, there was a closely related work~\cite{orito2024measurement} (arXiv: 2403.13435), which studies two particular diagonal cuts in the corresponding toric code in external fields via MOCs and their results agree with ours in where they overlap. 
}

\section*{Acknowledgement}
H.S. would like to thank Wenhan Guo for useful conversations regarding numerical simulations. 
He would also like to thank Takuya Okuda and Aswin Parayil Mana for discussions on related topics. 
This work was supported by the U.S. Department of Energy, Office of Science, National Quantum Information Science Research Centers, Co-design Center for Quantum Advantage (C2QA) under Contract No. DE-SC0012704 (KI, TCW). TCW acknowledges the support of Stony Brook University's  Center for Distributed Quantum Processing and the support by the National Science Foundation under Award No. PHY 2310614, in particular, on the connection between the MOC model with the measurement-based quantum computation.

\appendix

\section{Stabilizer update algorithm} \label{sec:stabilizer-update}
The algorithm for updating stabilizers can be found in Ref.~\cite{aaronson2004improved}, for example. 
Consider a state uniquely specified by stabilizers $S= \langle g_1,...,g_p\rangle$ and measuring an operator $\tilde{g}$.
(That is, $\tilde{g} = B_p$ in the measurement (1-1), for example.)
\begin{enumerate}
\item[(1)] Search in $\{g_1,...,g_p\}$ an element that anti-commutes with $\tilde{g}$. Let $g_k$ be that operator and go to 3.
\item[(2)] This is the case when $\tilde{g}$ commutes with all the elements in $S$. If $\tilde{g}$ or $-\tilde{g}$ is an element in $S$, then the measurement outcome is deterministic and the resulting stabilizer set remains $S$. Otherwise, update the stabilizer set to $\{g_1,...,g_p, (-1)^{m} \tilde{g}\}$, where $m=0,1$ is the measurement outcome. When the size of the stabilizer generators is the same as the number of qubits, then the second operation does not take place since the number of stabilizer should not increase.
\item[(3)] Swap $g_1$ and $g_k$.
\item[(4)] For $j=2,...,p$, check if $\tilde{g}$ anti-commutes with $g_i$, and if it does, update $g_i \mapsto g_1 g_i$. (After this, $g_1$ is the only operator non-commuting with $\tilde{g}$.)
\item[(5)] Measure $\tilde{g}$ with a random outcome $m$. (Both outcomes appear with an equality  probability of $1/2$.) Update the stabilizer set to $\{(-1)^m \tilde{g}, g_2,...,g_p\}$.
\end{enumerate}

\vspace{10pt}
\noindent
{\bf Example.}
If we start with a four-edge, four-vertex product state on a square, we have $S=\langle X_1, X_2, X_3, X_4, \tilde{X}_1, \tilde{X}_2, \tilde{X}_3, \tilde{X}_4\rangle$. We have represented the edge d.o.f. with tilde.
Consider measuring the plaquette operator $\tilde{g} = \tilde{Z}_1\tilde{Z}_2\tilde{Z}_3\tilde{Z}_4$. 
Then the algorithm goes as follows.
(1) We find an anti-commuting element $\tilde{X}_1$, $\{\tilde{X}_1,\tilde{g}\}=0$,  i.e., ``$g_k=\tilde{X}_1$."
(3) Swap $g_1=X_1$ with $g_k=\tilde{X}_1$. We have $S=\langle \tilde{X}_1, X_2, X_3, X_4,  X_1, \tilde{X}_2, \tilde{X}_3, \tilde{X}_4\rangle$. 
(4) In the right of the first element, we find $\tilde{X}_2, \tilde{X}_3, \tilde{X}_4$ anti-commute with $\tilde{g}=\tilde{Z}_1\tilde{Z}_2\tilde{Z}_3\tilde{Z}_4$.
We update as $S=\langle \tilde{X}_1, X_2, X_3, X_4,  X_1, \tilde{X}_1\tilde{X}_2, \tilde{X}_1\tilde{X}_3, \tilde{X}_1\tilde{X}_4\rangle$.
(5) Now we replace $g_1$ with $\tilde{g}$ due to the measurement, and all the other elements are unaffected since they commute with $\tilde{g}$: $S=\langle (-1)^m \tilde{Z}_1\tilde{Z}_2\tilde{Z}_3\tilde{Z}_4, X_2, X_3, X_4,  X_1, \tilde{X}_1\tilde{X}_2, \tilde{X}_1\tilde{X}_3, \tilde{X}_1\tilde{X}_4\rangle$, where $m$ denotes the measurement outcome.

\section{Entanglement entropy from stabilizers}\label{sec:EE-stabilizer}

To compute the entanglement entropy from stabilizers, we can employ the result by Hamma et al.~\cite{hamma2005bipartite}.
Here, we follow the description in Appendix C of Ref.~\cite{nahum2017quantum}.
Consider a subsystem $A$, and let $I_A$ be the number of independent stabilizers when restricted to the region $A$. 
The entanglement entropy is given by the following formula:
\begin{align}
S_A = I_A - |A| \ . 
\end{align}

Let us express the stabilizer using a $\mathbb{Z}_2 =\{0,1\}$ vector as
\begin{align}
S_j = X^{\vec{v}_j} Z^{\vec{w}_j} \ ,
\end{align}
with $\vec{v}_j, \vec{w}_j \in \{0,1\}^N$ being row vectors. 
We construct a vector 
\begin{align}
\vec{s}_j = (\vec{v}_j , \vec{w}_j) \ . 
\end{align}
Now we have a matrix at each time step
\begin{align}
\Psi(t) = ( \vec{s}^{ \, 1 T},\vec{s}^{\, 2 T}, ... , \vec{s}^{ \, |\mathcal{S}| T} ) \ .
\end{align}
Each column is a stabilizer and each row represents a Pauli.
Now $I_A$ is equivalent to the rank of the matrix $\Psi_A$ after deleting rows corresponding to $\bar{A}$. 
\begin{align}
I_A = \text{rank}( \Psi(t)_A ) \ .
\end{align}

\section{Pure-gauge MOC by measuring the RBH model}\label{sec:RBH}

In this Appendix, we consider the pure-gauge MOC circuit and relate it to the 3d surface percolation via a perspective of Measurement-Based Quantum Computation (MBQC)~\cite{raussendorf2001one,briegel2009measurement} on the Raussendorf-Bravyi-Harrington (RBH) state~\cite{raussendorf2005long}.  
We consider the 3d cubic lattice, and place qubits in the eigenstate $|+\rangle$ on faces and edges of it. 
We adopt a notation where the bold font denotes cells in the 3d lattice, while the ordinary font represents a cell in the $xy$-plane projected to the $z$ direction. 
For instance, we write $\mathbf{f} = f \otimes \{z=0\}$ for a face within a $xy$-plane at $z=0$, and  $\mathbf{f} = e \otimes [0,1]$ for a face extending in the $z$ direction;
we also write $\mathbf{e} = e \otimes \{z=0\}$ for an edge within a $xy$-plane, and  $\mathbf{e} = v \otimes [0,1]$ for an edge extending in the $z$ direction.
Then the RBH state is defined as 
\begin{align}
|\psi_\text{RBH}\rangle \equiv \prod_{ \langle \mathbf{e},\mathbf{f}\rangle } CZ_{\mathbf{e},\mathbf{f}} |+\rangle^{\otimes \mathbf{E} }  |+\rangle^{\otimes \mathbf{F} }. 
\end{align}
More precisely, we consider an open lattice with the range $z \in [0,L_z]$ and remove qubits on faces in the slice $z=L_z$.

We claim that the following measurements on the RBH state simulate the pure-gauge MOC model:
\begin{itemize}
    \item Measure each of the $\mathbf{E}$ qubits in the $X$ basis with the unit probability.
    \item Measure each of the $\mathbf{F}$ qubits in the $X$ basis with the probability $p_K$; otherwise measure the qubit in the $Z$ basis.
    \item The $\mathbf{E}$ qubits at $z=L_z$ remain unmeasured, where the simulated state is induced.
\end{itemize}
As in the main text, we can restrict our attention to the quantum trajectory with all the measurement outcomes being $+1$.

To see the correspondence, let us consider the RBH state with qubits in $z<\ell \leq L_z$ having been already measured. 
We can write 
\begin{align}
&|\psi_\text{RBH}(\ell) \rangle \nonumber \\ 
&= 
\prod_{ \langle \mathbf{e},\mathbf{f}\rangle , z\geq \ell } CZ_{\mathbf{e},\mathbf{f}} 
|+\rangle^{\otimes \mathbf{E}}_{ z>\ell}
|+\rangle^{\otimes \mathbf{F}}_{ z\ge\ell}
\otimes
|\Phi(\ell)\rangle_{\mathbf{E},z=\ell}
\otimes |\text{junk}\rangle ,
\end{align}
where $|\text{junk}\rangle$ is the normalized post-measurement product state in $z<\ell$ and $|\Phi(\ell)\rangle$ is the state on which we simulate the pure-gauge MOC.
It satisfies the condition $A_v|\Phi(\ell)\rangle=|\Phi(\ell)\rangle$.

We set the bra state associated with the measurement within $\ell \leq z<\ell +1$ to be
\begin{align}
\mathbb{M}_{\ell \leq z<\ell +1}
= \Big(
\prod_{\mathbf{e} \in \mathbf{E}}
\langle + | 
\prod_{\mathbf{f} \in \mathbf{F}_X }
\langle + |
\prod_{\mathbf{f} \in \mathbf{F}_Z}
\langle 0 | \Big) \quad \Big|_{\ell \leq z<\ell+1} .
\end{align}
Here $\mathbf{F}_X$ ($\mathbf{F}_Z$) denotes the subset of faces where the measurement with the $X$ ($Z$) basis occurred.
Then the state $\mathbb{M}_{\ell \leq z<\ell +1}|\psi_\text{RBH}(\ell) \rangle$ is equal to the state $|\psi_\text{RBH}(\ell+1) \rangle$ with the simulated state having evolved as
\begin{align}
&|\Phi(\ell+1)\rangle
=
\prod_{ e \times [\ell,\ell+1] \in \mathbf{F}_Z} |+\rangle \langle 0|_e 
\prod_{ e \times [\ell,\ell+1] \in \mathbf{F}_X} \frac{H_e}{\sqrt{2}} \nonumber \\
& \qquad \qquad
\times \prod_{v \in V} \frac{1+\prod_{e \supset v} Z_e }{2}
\prod_{e \in E} \frac{H_e}{\sqrt{2}} \nonumber \\
&\qquad \qquad \times \prod_{f \times \{\ell\} \in \mathbf{F}_Z} \frac{1}{\sqrt{2}} \prod_{f \times \{\ell\} \in \mathbf{F}_X} \frac{1+B_f}{2}
|\Phi(\ell)\rangle \nonumber \\
&=  \prod_{e \in E} \frac{1}{\sqrt{2}} \times 
\prod_{ e \times [\ell,\ell+1] \in \mathbf{F}_Z} |+\rangle \langle +|_e 
\prod_{ e \times [\ell,\ell+1] \in \mathbf{F}_X} \frac{1}{\sqrt{2}} \nonumber \\
&\times \prod_{f \times \{\ell\} \in \mathbf{F}_Z} \frac{1}{\sqrt{2}} \prod_{f \times \{\ell\} \in \mathbf{F}_X} \frac{1+B_f}{2}
|\Phi(\ell)\rangle . 
\end{align}
Thus one obtains a model where the measurement with $B_f$ occurs with probability $p_K$ and that with $X_e$ with probability $1-p_K$. 
We denote the entire measurement sequence as $\mathbb{M} = \prod_{\ell \in \{0,...,L_z-1\}} \mathbb{M}_{\ell \leq z<\ell +1} $, so that the simulated state in the end is $\mathbb{M}|\psi_\text{RBH}\rangle$.

We mentioned that the simulated state obeys the Gauss law constraint $A_v=+1$. 
The long-range order would be supported in coordination with the 
vacuum expectation value with respect to a non-local operator, i.e.,
$\langle W(\gamma) \rangle$ with $W(\gamma)$ a product of $Z$ operators along an arbitrarily large closed loop.
We note that the RBH state is symmetric under a transformation supported on relative cycles $\mathbf{z}$, which is a sum of elements in $\mathbf{F}$ such that its boundary is a closed loop on $z=L_z$ (a closed surface in 3d which terminates on the boundary as a loop), i.e., $\partial \mathbf{z} =\gamma$:
\begin{align}
&\mathbf{W}(\mathbf{z}) \equiv \prod_{\mathbf{f} \in \mathbf{z} }X_{\mathbf{f}} \times W(\gamma) , \\
& \mathbf{W}(\mathbf{z})  |\psi_\text{RBH}\rangle =  |\psi_\text{RBH}\rangle, 
\end{align}
where, in the second equality, we have used that $\mathbf{W}(\mathbf{z})$ is a product of stabilizer operators of the RBH graph state.
The condition is also known as the 1-form symmetry of the RBH state, see e.g., Ref.~\cite{roberts2017symmetry}.
Hence, if a subset of $\mathbf{F}_X$ in our measurement forms relative cycles, it enforces correlation in the simulated state.
Namely, if $\mathbf{z}$ can be formed by the cells in $\mathbf{F}_X$, then
\begin{align} \label{eq:bulk-boundary-symm-cond}
\mathbb{M} \mathbf{W}(\mathbf{z})  |\psi_\text{RBH}\rangle  = 
\mathbb{M} W(\gamma) |\psi_\text{RBH}\rangle  = \mathbb{M} |\psi_\text{RBH}\rangle ,
\end{align}
so that $W(\gamma)$ becomes 
enforced in
the simulated state.

Our equation~\eqref{eq:bulk-boundary-symm-cond} implies that if faces in $\mathbf{F}_X$ can form large cycles (closed surfaces, except those that terminate at the boundary), it is likely that the 2d simulated state hosts the deconfining order supported by large Wilson loops ($W(\gamma) =1$ for a large $\gamma$).
Roughly speaking, the formation of giant cycles (closed surface consisting of faces) by cells each of which is activated with probability $p_K$ is the mathematical problem called homological percolation.
One of its simplest versions in lower dimensions, e.g. in 2d, is the bond percolation problem, which is the underlying mechanism of the MIPT in the pTFI model~\cite{2020PhRvB.102i4204L}.
The spacetime picture we provided here is a generalization of theirs, phrased in the language of MBQC.
We speculate that the transition to the deconfining phase occurs as the consequence of the homological percolation formed by $X$ measurements on the RBH state.

\rev{In Ref.~\cite{Okuda-Parayil-Mana-Sukeno}, a generalization of the RBH model was constructed in the context of quantum simulation of (unitary) real-time dynamics in the Fradkin-Shenker model. 
Analogously to the discussion in this Appendix, one can generate the FS-MOC in the main text by performing single-qubit projective measurements on the resource state in Ref.~\cite{Okuda-Parayil-Mana-Sukeno}, where the choice between the $X$ and $Z$ basis is randomized.
}

\begin{figure}[h]
\includegraphics[width=0.8\linewidth]{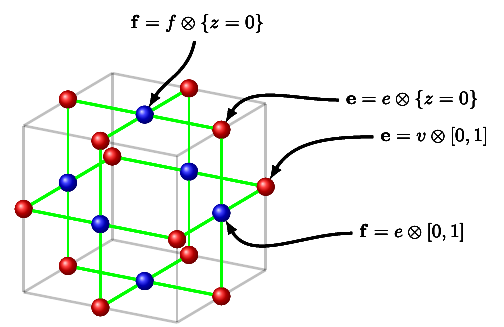}
\caption{The unit cell in the RBH state. Qubits are placed on the edges and faces of the 3d cubic lattice, and $CZ$ gates are applied to adjacent qubits as drawn in green. Note that we place $z=0$ at the top and the simulated  time direction goes from the top to bottom.}
\label{fig:RBH}
\end{figure}

\rev{
\section{$ZX$-randomized Fradkin-Shenker MOC}
}
\label{sec:XZ-randomized}

\rev{
In the FS-MOC model in the main text, we consider  measurement circuits alternating between the $Z$-bases measurements and the $X$-bases measurements.
As indicated by our numerical result, the entanglement measures and the Wilson operator expectation values exhibit different behaviors depending on whether we evaluate them after the $Z$-type measurements or after the $X$-type measurements.
An alternative MOC model that resolves this dependency would be an {\it $ZX$-randomized Fradkin-Shenker MOC}.
(See e.g. Refs.~\cite{lavasani2021topological, zhu2023structured} for related setups.)
Namely, one can introduce another independent probability $0 \leq p_{ZX} \leq 1$, and at each round the $Z$ bases measurements take place with probability $p_Z$; otherwise, the $X$ bases measurements are performed. 
Here, we investigate such a model with $p_{ZX} =\frac{1}{2}$; the detail of the phase diagram would change as we vary $p_{ZX}$, but simulation over varying $p_{ZX}$ is beyond scope of our study in this work.
We present a pseudo-code in Algorithm~\ref{alg:XZ-randomized}. 
}

\begin{figure}
\begin{minipage}{\linewidth}
\begin{algorithm}[H]
\begin{algorithmic}
    \For {$s=1,...,N_s$}
        \State Initialize $|+\rangle^V |+\rangle^E$.
        \For {$t=1,...,N_t$}
            \If { $\text{rand}()<p_{ZX}$ }
                \State $Z$-bases measurement round. (The $W_e$ term with probability $p_J$, and the $B_p$ term with probability $p_K$)
            \Else
                \State $X$-bases measurement round. (The $X_v$ term with probability $1-p_J$, and the $X_e$ term with probability $1-p_K$)
            \EndIf 
        \EndFor
        \State Compute physical quantities (entanglement measures and the Wilson line operator).
    \EndFor
    \State Compute the average of physical quantities.
\end{algorithmic}
\caption{$ZX$-randomized Fradkin-Shenker MOC}
\label{alg:XZ-randomized}
\end{algorithm}
\end{minipage}
\end{figure}

\begin{figure*}
    \includegraphics[width=\linewidth]{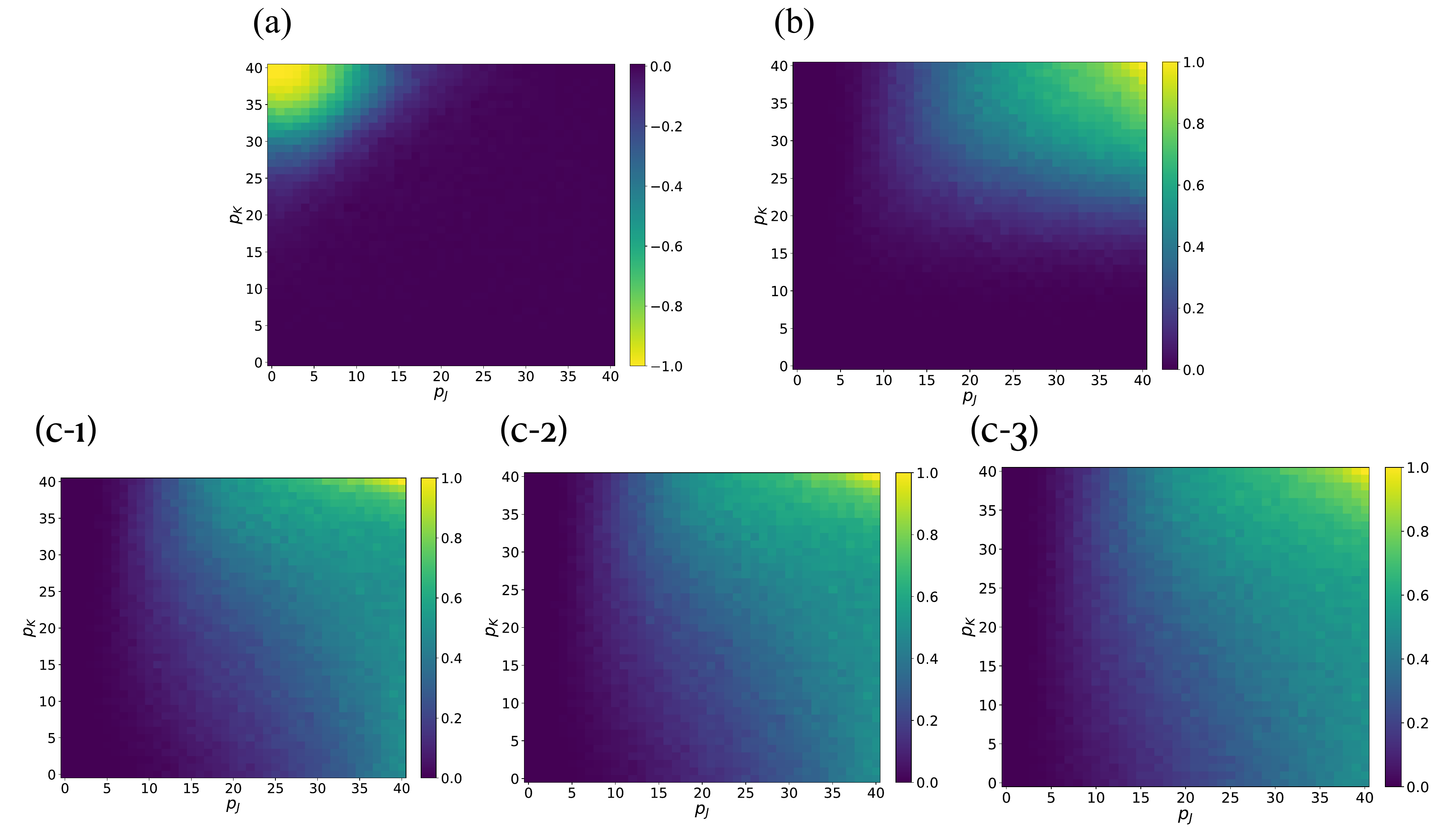}
    \caption{Numerical results for the $ZX$-randomized Fradkin-Shenker MOC with $(L_x,L_y,N_s,N_t) =(12,7,1000,30)$. (a) The topological entanglement entropy. (b) The boundary mutual information. (c) The open Wilson line operator. Length of Wilson lines: (c-1) 7, (c-2) 5, (c-3) 3. }
    \label{fig:rand-result-1}
\end{figure*}

\begin{figure*}
    \includegraphics[width=\linewidth]{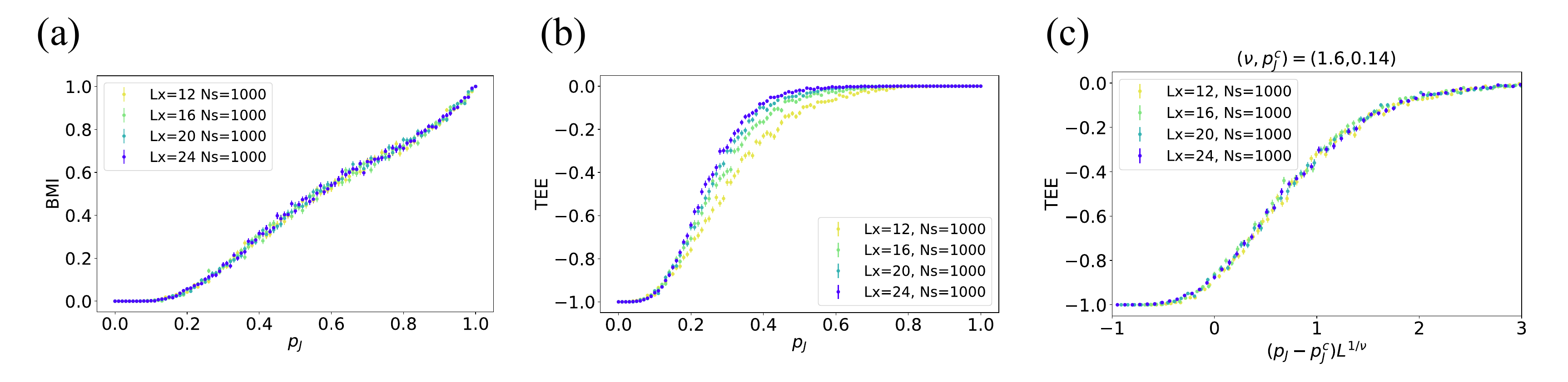}
    \caption{The $ZX$-randomized Fradkin-Shenker MOC scanned at $p_K=1.0$. (a) the BMI showing a transition at $p_J=0.12(2)$, (b) the TEE, and (c) a finite-size scaling analysis for the TEE indicates the transition at $p_c=0.14 (2)$ with critical exponent $\nu = 1.6 (2)$.}
    \label{fig:rand-result-2}
\end{figure*}

\rev{
We present data of our numerics with $N_s=1000$ and $N_t=30$ in Fig.~\ref{fig:rand-result-1} and in Fig.~\ref{fig:rand-result-2}.
We summarize some observation below, and conclude that the essential feature is consistent with the recent proposition of Higgs=SPT.
}

\rev{
The phase diagram charted in Fig.~\ref{fig:rand-result-1}(a) indicates that the deconfining phase detected by the topological entanglement entropy qualitatively matches that of the original FS-MOC model. 
However, details of the phase diagram changes; e.g., the non-zero TEE is exhibited in a smaller region and the cusp is less prominent.
For example, the scan given in Fig.~\ref{fig:rand-result-2}(b) and the finite size scaling in Fig.~\ref{fig:rand-result-2}(c) indicate a transition along $p_K=1$ at $p_c=0.14 (2)$ with critical exponent $\nu = 1.6 (2)$. 
}

\rev{
The phase diagram charted in Fig.~\ref{fig:rand-result-1}(b) indicates that the Higgs=SPT phase detected also qualitatively matches the original FS-MOC model. 
However, it is exhibited over a larger region as indicated by Fig.~\ref{fig:rand-result-2}(a), which shows that the transition along $p_K=1$ take place at $p_J=0.12(2)$. 
The four different lattice sizes show the same curve to the precision of our numerics.
}

\rev{
The phase diagram charted in Fig.~\ref{fig:rand-result-1}(c) shows an averaged version of the mixed phases as exhibited by the open Wilson line operators.
The longer the line is, slightly sharper the non-zero region becomes.
}

\rev{\section{Entanglement measures after a half cycle}}
\label{sec:half}

\begin{figure*}
    \centering
    \includegraphics[width=\linewidth]{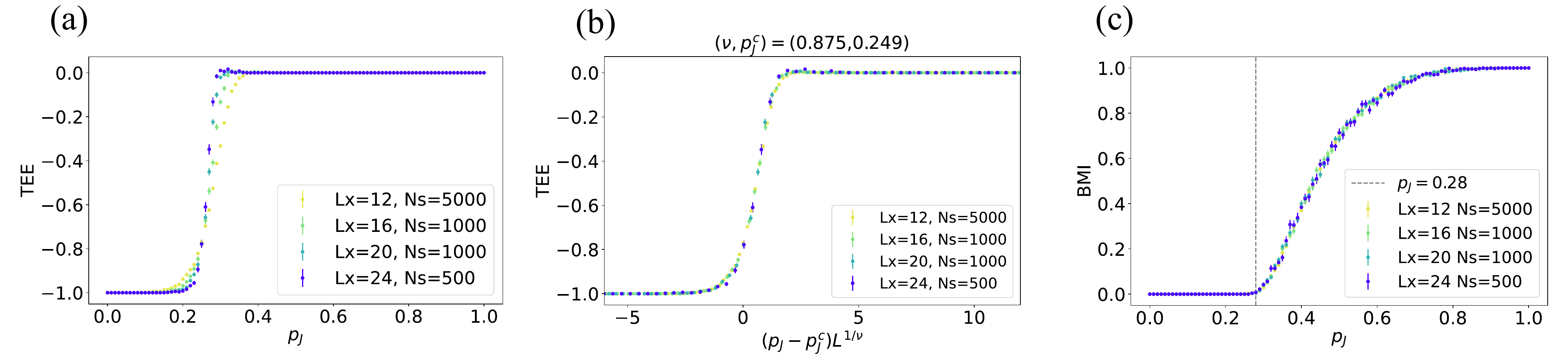}
    \caption{
    \rev{Numerical results along the scan (i) $p_K=1$ at the half cycle after 30 full cycles. (a) The TEE. (b) The finite size scaling analysis for the TEE, indicating the transition at $p^c_J = 0.25(1)$ with critical exponent $\nu = 0.87(5)$. (c) The BMI, indicating the transition at $p^c_J = 0.28(2)$.}
    }
    \label{fig:cut_i_half}
\end{figure*}

\rev{In Fig.~\ref{fig:cut_i_half}, we provide data on simulations where we computed entanglement measures after a half cycle after $N_t=30$ full cycles.
We examine the scan (i) along the $p_K=1$ submanifold. 
We find that Fig.~\ref{fig:cut_i_half}(b) indicates a phase transition at $p^c_J=0.25(1)$ with critical exponent $\nu = 0. 87(5)$, which is consistent with the 3d bond percolation as in the full cycle case. 
The BMI in Fig.~\ref{fig:cut_i_half}(c) also shows that the threshold probability is consistent with the full cycle case. }

\bibliography{refs}
\end{document}